%% file: main_fase.tex
\documentclass[runningheads]{llncs}
\usepackage{graphicx}
\usepackage[noadjust]{cite}

\usepackage{multirow}
\usepackage{amssymb}
\usepackage{graphicx}
\usepackage{color}
\usepackage{fancybox}
\usepackage{xcolor}
\usepackage{booktabs}
\usepackage{framed}
\usepackage{colortbl}
\usepackage{xspace}
\usepackage{enumitem}
\usepackage{verbatim}
\usepackage{siunitx}

\usepackage[turnon]{notes}
\usepackage[switch]{lineno}
\usepackage{lineno}
\usepackage{amsmath}
\usepackage{blkarray, bigstrut}
\usepackage{wrapfig}

\makeatletter
\RequirePackage[bookmarks,unicode,colorlinks=true]{hyperref}%
   \def\@citecolor{blue}%
   \def\@urlcolor{blue}%
   \def\@linkcolor{blue}%

\def\orcidID#1{\smash{\href{http://orcid.org/#1}{\protect\raisebox{-1.25pt}{\protect\includegraphics{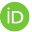}}}}}
\makeatother

\usepackage{url}
\include{includes}
\include{macros}

\usepackage{cleveref}
\usepackage[T1]{fontenc}

\begin{document}
\title{Understanding Local Robustness of Deep Neural Networks under Natural Variations}
\titlerunning{Understanding Local Robustness of DNNs under Natural Variations}
\author{Ziyuan Zhong\orcidID{0000-0001-9661-1233} \and
Yuchi Tian\orcidID{0000-0002-9711-1449} \and
Baishakhi Ray\orcidID{0000-0003-3406-5235}}

\authorrunning{Z. Zhong et al.}

\institute{Columbia University, New York, NY, USA
\email{\{ziyuan.zhong, yuchi.tian\}@columbia.edu, rayb@cs.columbia.edu}}
\maketitle              % typeset the header of the contribution
\begin{abstract}
Deep Neural Networks (DNNs) are being deployed in a wide range of settings today, from safety-critical applications like autonomous driving to commercial applications involving image classifications. However, recent research has shown that DNNs can be brittle to even slight variations of the input data. Therefore, rigorous testing of DNNs has gained widespread attention.

While DNN robustness under norm-bound perturbation got significant attention over the past few years, our knowledge is still limited when natural variants of the input images come.
These natural variants, \eg a rotated or a rainy version of the original input,  are especially concerning as they can occur naturally in the field without any active adversary and may lead to undesirable consequences.
Thus,  it is important to identify the inputs whose small variations may lead to erroneous DNN behaviors. 
The very few studies that looked at DNN's robustness under natural variants,  however, focus on estimating the overall robustness of DNNs across all the test data rather than localizing such error-producing points. This work aims to bridge this gap.

To this end, we study the local per-input robustness properties of the DNNs and leverage those properties to build a white-box (\toolWB) and a black-box (\toolBB) tool to automatically identify the non-robust points.  Our evaluation of these methods on three DNN models spanning three widely used image classification datasets shows that they are effective in flagging points of poor robustness. In particular, \toolWB and \toolBB  are able to achieve an F1 score of up to 91.4\% and 99.1\%, respectively. We further show that \toolWB can be applied to a regression problem in a domain beyond image classification. Our evaluation on three self-driving car models demonstrates that \toolWB is effective in identifying points of poor robustness with F1 score up to 78.9\%.

\end{abstract}
%
%
%

\input{introduction}
\input{background}

\input{theory.tex}

\input{methodology}
\input{experiments.tex}
\input{rq2}
\input{rq3}
\input{rq4}

\input{rq5}
\input{related_work}
\input{threats}
\input{conclusion}
\input{acknowledgement}

\bibliographystyle{splncs04}
\bibliography{./references.bib}
\newpage
\appendix
\input{appendix}

\end{document}

%% file: includes.tex
\usepackage{multirow}
\usepackage{amssymb}
\usepackage{graphicx}
\usepackage{color}
\usepackage{fancybox}
\usepackage{xcolor}
\usepackage{booktabs}
\usepackage{framed}
\usepackage{colortbl}
\usepackage{xspace}
\usepackage{enumitem}
\usepackage{verbatim}
\usepackage{siunitx}

\usepackage{cleveref}
\usepackage[turnon]{notes}
\usepackage{amsmath}
\usepackage{blkarray, bigstrut}
\usepackage{balance}
\usepackage[normalem]{ulem}

\usepackage[font=small,skip=0pt]{caption}
\usepackage{tabularx}

%% file: macros.tex
\usepackage{framed}
\usepackage{enumitem}
\usepackage{subfig}
\usepackage{mdframed}

\newcommand{\ie}{i.e.,\xspace}
\newcommand{\eg}{e.g.,\xspace}
\newcommand{\etc}{etc.\xspace}
\newcommand{\wrt}{\textit{w.r.t.}\xspace}
\newcommand{\etal}{\textit{et al.}\xspace}
\newcommand{\aka}{\textit{a.k.a.}\xspace}

\newcommand{\resnet}{ResN}
\newcommand{\wrn}{WRN}
\newcommand{\vgg}{VGG}
\newcommand{\cifar}{CIFAR-10}
\newcommand{\sdc}{Self-Driving}
\newcommand{\svhn}{SVHN}
\newcommand{\fmnist}{F-MNIST}
\newcommand{\celebA}{CELEB-A}
\newcommand{\toolname}{\textsc{Deep\-Robust}\xspace}

\newcommand{\toolBB}{\textsc{Deep\-Robust-B}\xspace} % Black-box analysis
\newcommand{\toolWB}{\textsc{Deep\-Robust-W}\xspace} % White-box analysis
\newcommand{\mut}{{DNN under test}\xspace}

\newcommand{\nbc}[3]{
	{\colorbox{#3}{\bfseries\sffamily\scriptsize\textcolor{white}{#1}}}
	{\textcolor{#3}{\sf\small$\blacktriangleright$\textit{#2}$\blacktriangleleft$}}
}
\definecolor{todocolor}{rgb}{1,0.3,0.3}

\newcommand{\cmt}[1]{\nbc{COMMENT}{#1}{todocolor}}

\newcommand{\edited}[1]{{\color{black} #1}}

\newcommand{\RQ}[2]{%
 \noindent
 \textbf{\large RQ#1.~#2}
}

\newcommand{\RQA}[2]{%
\noindent
\uline{RQ#1.~#2}
}
\definecolor{shadecolor}{rgb}{179,177,170}

\definecolor{gray05}{gray}{0.95}
\definecolor{gray10}{gray}{0.90}
\definecolor{gray12}{gray}{0.88}
\definecolor{gray15}{gray}{0.85}
\definecolor{gray20}{gray}{0.80}
\definecolor{gray30}{gray}{0.70}
\definecolor{gray40}{gray}{0.60}
\definecolor{gray50}{gray}{0.50}
\definecolor{gray60}{gray}{0.40}
\definecolor{gray70}{gray}{0.30}
\definecolor{gray80}{gray}{0.20}
\definecolor{gray90}{gray}{0.10}
\definecolor{fig1_blue}{RGB}{85, 85, 255}
\definecolor{fig1_green}{RGB}{34, 148, 34}
\definecolor{fig1_gray}{gray}{0.11}
\definecolor{lightgray}{gray}{0.8}
\definecolor{darkgray}{gray}{0.6}
\definecolor{Gray}{rgb}{0.88,1,1}
\definecolor{Gray}{gray}{0.85}
\definecolor{Blue}{RGB}{0,29,193}
\definecolor{MyDarkBlue}{rgb}{0,0.08,0.45} 
\definecolor{pink}{RGB}{231,95,110}
\definecolor{lightestgray}{rgb}{0.95, 0.95, 0.95}

\newcommand{\RS}[2]{
{
\begin{mdframed}[backgroundcolor=gray15, 
  topline=false,rightline=false,leftline=false, bottomline=false] 
\filbreak
\noindent
\textbf{Result #1:}~\emph{#2}
\end{mdframed}
}
}

\newcommand*\median[1]{\hat{#1}}

\newcommand{\rqb}{What are the characteristics of the weak  points?\xspace}
\newcommand{\rqc}{Can we detect the weak points in a white-box setting?\xspace}
\newcommand{\rqd}{Can we identify the weak points in a black-box setting?\xspace}
\newcommand{\rqe}{How generalizable are these findings?\xspace}

\makeatletter
\newcommand*{\inlineequation}[2][]{%
  \begingroup
    \refstepcounter{equation}%
    \ifx\\#1\\%
    \else
      \label{#1}%
    \fi
    \relpenalty=10000 %
    \binoppenalty=10000 %
    \ensuremath{%
      #2%
    }%
    ~\@eqnnum
  \endgroup
}
\makeatother

%% file: introduction.tex
%auto-ignore
\section{Introduction}\label{sec:intro}

Deep Neural Networks (DNNs) have achieved an unprecedented level of performance over the last decade in many sophisticated areas such as image recognition~\cite{krizhevsky2012imagenet}, self-driving cars~\cite{bojarski2016end} and playing complex games~\cite{44806}. 
These advances have also motivated companies to 
%integrate AI capabilities into software and services, and to 
adapt their software development flows to incorporate AI components~\cite{MSR-ICSE-SEIP.2019}. 
This trend has, in turn, spawned a new area of research within software engineering addressing the quality assurance of DNN components~\cite{pei2017deepxplore,tian2017deeptest,zhang2018deeproad, ma2018mode, kim2019guiding, CRADLE:ICSE2019, DeepStellar:FSE2019, Islam:FSE2019, Li:FSE2019, Apricot:ASE2019, tian2019deepinspect, Sensei:ICSE2020}. 

Notwithstanding the impressive capabilities of DNNs, recent research has shown that DNNs can be easily fooled, \ie made to mispredict, with a little variation of the input data~\cite{goodfellow2014explaining,engstrom2019exploring,tian2017deeptest}\textemdash either adding a norm-bound pixel-level perturbation into the original input~\cite{szegedy2013intriguing,goodfellow2014explaining,carlini2017towards}, or with \emph{natural} variants of the inputs, \eg rotating an image, changing the lighting conditions, \edited{adding fog} \etc~\cite{engstrom2019exploring, pei2017deepxplore, Ozdag2019OnTS}. 
The natural variants are especially concerning as they can occur naturally in the field without any active adversary and may lead to serious consequences~\cite{tian2017deeptest,zhang2018deeproad}.

\begin{figure}[t]
\centering
    \subfloat[\ang{0}, bird\label{fig:bird1}]
    {\includegraphics[width=0.11\textwidth]{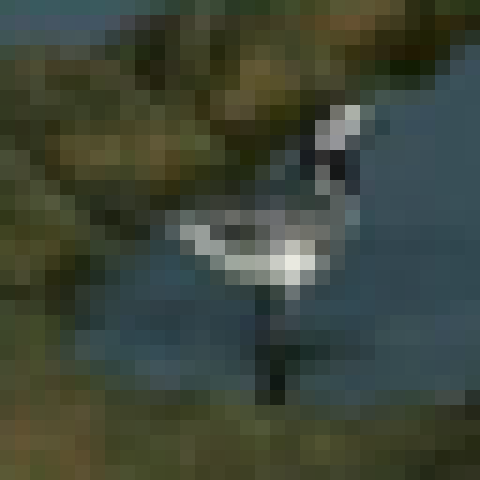}}
    \hspace{0.01\textwidth}%
    \subfloat[+\ang{6}, airplane]
    {\includegraphics[width=0.11\textwidth]{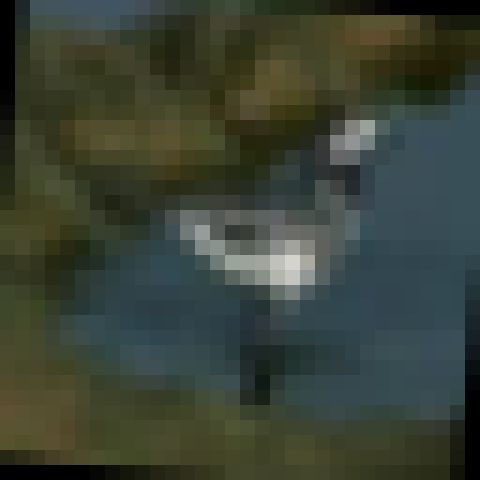}}
    \hspace{0.01\textwidth}%
    \subfloat[+\ang{24}, cat]
    {\includegraphics[width=0.11\textwidth]{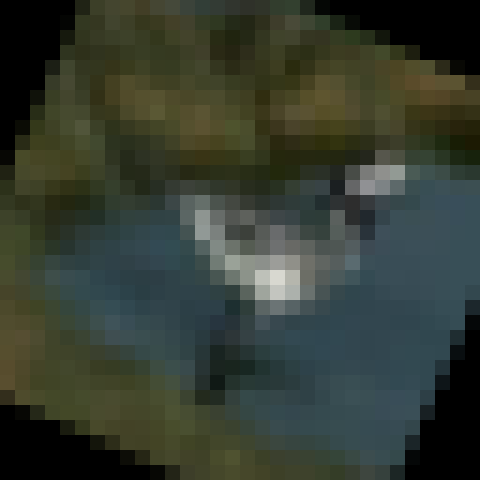}}
    \hspace{0.01\textwidth}%
    \subfloat[-\ang{9}, dog]
    {\includegraphics[width=0.11\textwidth]{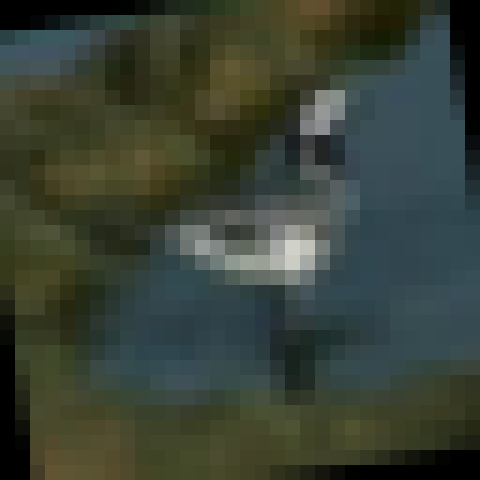}}
    \hspace{0.02\textwidth}%
    \subfloat[\ang{0}, bird\label{fig:bird2}]
    {\includegraphics[width=0.11\textwidth]{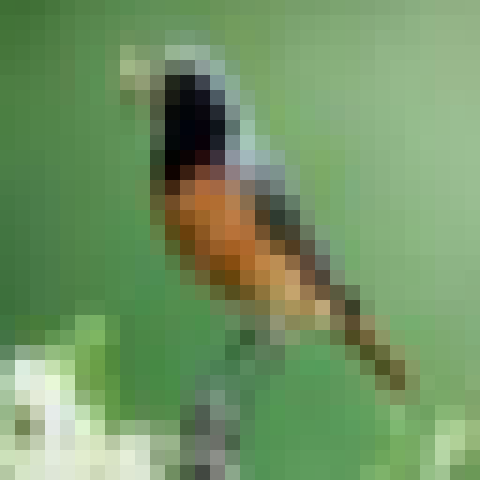}}
    \hspace{0.01\textwidth}%
    \subfloat[+\ang{6}, bird]
    {\includegraphics[width=0.11\textwidth]{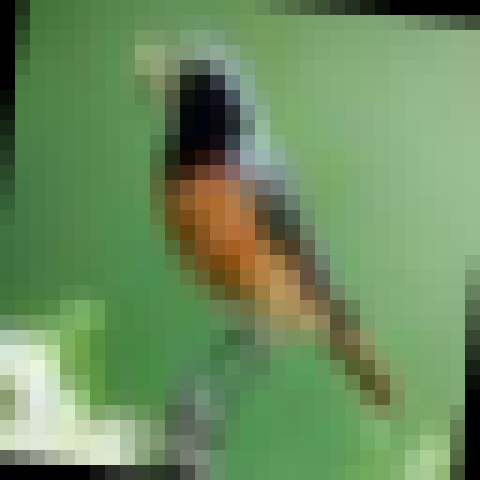}}
    \hspace{0.01\textwidth}%
    \subfloat[+\ang{24}, bird]
    {\includegraphics[width=0.11\textwidth]{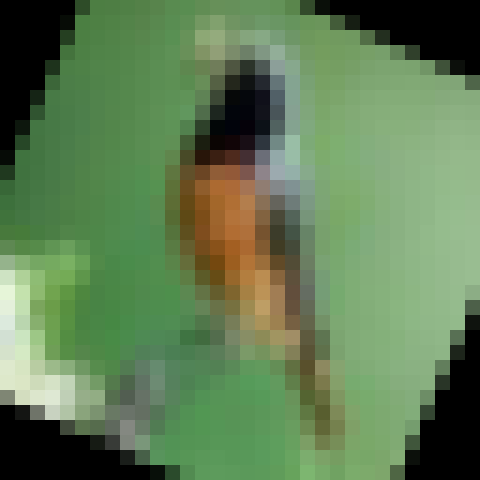}}
    \hspace{0.01\textwidth}%
    \subfloat[-\ang{9}, bird]
    {\includegraphics[width=0.11\textwidth]{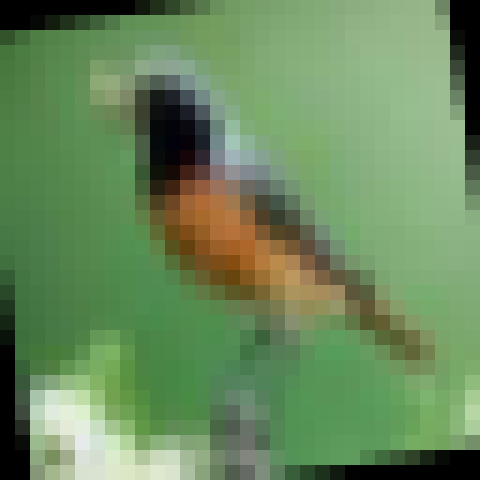}}
\caption{\textbf{\small{\textbf{(a)-(d)} A well-trained Resnet model~\cite{engstrom2019exploring} misclassifies the rotated variations of a bird image into three different classes though the original un-rotated image is classified correctly. 
\textbf{(e)-(h)} The same model successfully classifies all the rotated variants of another bird image from the same test set. The sub-captions consist of rotation degrees and the predicted classes.}}}
\label{fig:motivation_different_classes}
\vspace{-8mm}
\end{figure} 

While norm-bound perturbation based DNN robustness is relatively well-studied, 
our knowledge of DNN robustness under the natural variations is still
limited\textemdash we do not know which images are more robust than others, what their characteristics are, etc. 
For example, consider~\Cref{fig:motivation_different_classes}: although the original bird image (a) is predicted correctly by a DNN, its rotated variations in images (b)-(d) are mispredicted to three different classes. This makes the original image (a) very weak as far as robustness is concerned.  In contrast, the bird image (e) and all its rotated versions (generated by the same degrees of rotation) in 
Figure~\ref{fig:motivation_different_classes}:(f)-(h) are correctly classified. Thus, the original image (e) is quite robust.  
It is important to distinguish between such robust vs.~non-robust  images, as the non-robust ones can induce errors with slight natural variations.

Existing literature, however, focuses on estimating the overall robustness of DNNs across all the test data ~\cite{engstrom2019exploring, trade_alp_spatial,geometric_certify}. From a traditional software point of view, this is analogous to estimating how buggy a software is without actually localizing the bugs.  Our current work tries to bridge this gap by localizing the non-robust points in the input space that pose significant threats to a DNN model's robustness. However, unlike traditional software where bug localization is performed in program space,  we identify the non-robust inputs in the data space. As a DNN is a combination of data and architecture, and the architecture is largely uninterpretable, we restrict our study of non-robustess to the input space. 
To this end, we first quantify the local (per input) robustness 
property of a DNN. 
First, we treat all the natural variants of an input image as its {\em neighbors}. 
Then, for each input data, we consider a population of its neighbors and measure the fraction of this population classified correctly by the DNN - a high fraction of correct classifications indicates good robustness (\Cref{fig:motivation_different_classes}:e) and vice versa (\Cref{fig:motivation_different_classes}:a). We term this measure {\em neighbor accuracy}. Using this metric, we study different local robustness properties of the DNNs and analyze how
the weak, \aka non-robust, points differ characteristically from their robust counterparts. Given that the number of natural neighbors of an image can be potentially infinite, 
first we performed a more controlled analysis by keeping the natural variants limited to spatially transformed images generated by rotation and translation, following the previous work~\cite{engstrom2019exploring,trade_alp_spatial,geometric_certify}.
Such controlled experiments help us to explore different robustness properties while systematically varying transformation parameters.%\baishakhi{"while" has grammar mistake}

Our analysis with three well-known object recognition datasets across three popular DNN models, \ie a total of nine DNN-dataset combinations, reveal several interesting properties of local robustness of a DNN \wrt natural variants:

\begin{itemize}[leftmargin=*,topsep=0pt]
% \item 
% Local robustness properties seem to be intrinsic to the dataset rather than the model, \ie most of the robust points remain robust across different models and vice versa. 

\item The neighbors of a weaker point are not necessarily classified to one single incorrect class. In fact, the weaker the point is its neighbors (mis)classifications become more diverse. 

\item The weak points are concentrated towards the class decision boundaries of the DNN in the feature space.
\end{itemize}

Based on these findings, we further develop two techniques (a black-box and a white-box) that can localize the points of poor robustness, %at runtime, 
thereby providing a means of, input-specific, real-time feedback about robustness to the end-user.
Our white-box and black-box detectors can identify weak, \aka non-robust, points with f1 score up to 91.4\% and 99.1\%, respectively, at neighbor accuracy cutoff 0.75. 
{To further check the generalizability of our technique, we aim to detect weak points \wrt a self-driving car application where we generated natural input variants by adding rain and fog. Note that these are more  complex image transformations, and also the model works in a regression setting instead of classification. These models take an image as input, and output a driving angle. Our white-box detector can identify weak points with f1 score up to 78.9\%.}

%%% Contributions
In summary, we make the following contributions: 
\begin{itemize}[leftmargin=*,topsep=0pt]
    \item %\textbf{Empirical Characterization:} 
    We conduct an empirical study to understand the local robustness properties of DNNs under natural variations.
    \item %\textbf{Detection of weak points:} 
    We develop a white-box (\toolWB) and a black-box (\toolBB) method to automatically detect weak points.% in real-time. 
    \item
    We present a detailed evaluation of our methods on three DNN models across three image classification datasets. To check the generalizability of our findings, we further evaluate \toolWB in a setting with non-spatial transformations (\ie rain and fog),  a different task (\ie regression), and a safety-critical application (\ie self-driving car).
    We find that \toolname can successfully detect weak points with reasonably good precision and recall. 
    \item
    We made our code public at \url{https://github.com/AIasd/DeepRobust}.  
    % We made our analysis script and code public at: \\
    % \url{https://github.com/deeprobust/DeepRobust}.  
\end{itemize}

%The rest of the paper is organized as follows.~\ziyuan{fill up}

%% file: background.tex
%auto-ignore
\section{Background: DNN Testing}
\label{sec:background}

Existing studies have proposed different techniques to generate test data inputs by perturbing input images for a DNN and use them to evaluate the robustness of the DNN. Depending on how the input image is perturbed, the techniques for generating DNN test data can be classified into three broad categories: 
%norm-based perturbation and image transformation based perturbation.

\textit{i) Adversarial inputs} are typically generated by norm-based perturbation techniques~\cite{goodfellow2014explaining,kurakin2016adversarial,papernot2016limitations,carlini2017towards,madry2017towards,xiao2018generating} where some pixels of an input image ($I$) are perturbed by norm-based distance ($l_1$,$l_2$ or $l_{\inf}$) such that the distance between the perturbed image and $I$ is $\leq \epsilon$, where $\epsilon$ is a small positive value. These adversarial examples are used to expose the security vulnerabilities of DNNs. 

\textit{ii) Natural variations} are generated 
%by naturally-occurring variations  in  the  configuration  or  context  of  data  capture, 
through a variety of image transformations, and are used to evaluate the robustness of DNNs under such variations~\cite{engstrom2019exploring,engstrom2017rotation,tian2017deeptest}. Sources of these variations include changes in camera configuration, or variations in background or ambient conditions. The transformations simulating these variations could be spatial, such as rotation, translations, mirroring, shear, and scaling on images, or non-spatial transformations, such as changes in the brightness or contrast %, or adding color balance 
of an image. 
% \ziyuan{write why do we focus on spatial transformation?}
Here we first focus on spatial transformations as opposed to adversarial one for two reasons. %\baishakhi{this is ambiguous. Spatial transformations vs non-spatial transformation or spatial transformation vs adversarial examples?}
First, compared with adversarial examples, which is fairly contrived, spatial transformations are more likely to arise in more benign environments. 
Second, using simple parametric spatial transformations like rotations and translations, it is easier to systematically explore the local robustness properties.
%although the vulnerability of state-of-the-art DNNs to spatial transformations have been found recently~\cite{engstrom2019exploring}, the properties of such vulnerability are less well-understood.
% ~\ziyuan{read the following:add fog rain}.
% Also, parametric spatial transformations help us to perform controlled experiments. 
Later, to emulate a more natural variation we add fog and rain
on the images of self-driving car dataset and evaluate our method's generalizibility. 
%Recent work has demonstrated that modern DNNs are particularly vulnerable to spatial transformations~\cite{engstrom2019exploring}. Therefore, our work also focuses on this class of natural variations. 
%Spatial transformation can be thought of as a mapping function between the location of all points in an original image and the location of all points in a new image. 
%There are quite a few work~\cite{engstrom2019exploring,engstrom2017rotation,tian2017deeptest} that have leveraged spatial transformations-based adversarial examples to evaluate the robustness of a DNN model.
%\subsubsection{Spatial transformation based}
% Spatial transformations based perturbation is to apply spatial transformation on images with a small parameter to generate adversarial examples. Spatial transformation is a mapping function between the location of all points in an original image and the location of all points in a new image. Spatial transformations include rotation, vertical/horizontal translation, mirroring, shear, scaling and affine transformations and so on. Some works\cite{engstrom2019exploring,engstrom2017rotation,tian2017deeptest} leverage spatial transformations to evaluate the robustness of a DNN model. In this work, we propose to apply spatial transformations (we select rotation as a representative of non-linear spatial transformation and translation as a representative of linear spatial transformation) on a data point to generate neighbors and leverage how a DNN classifies or predicts the neighbors of each point to measure the given DNN's robustness at the data point.

\textit{iii) GAN-based} image generation techniques use Generative
Adversarial Network (GAN) to synthesize images. GAN is one class of generative models trained as a minimax two-player game between a generative model and a discriminative model~\cite{goodfellow2014generative}. GAN-based image generation has been successfully used to generate DNN test data instances~\cite{zhang2018deeproad,zhao2017generating}.

\noindent
\textbf{Standard Accuracy vs. Robust Accuracy.} Standard accuracy measures how accurately an ML model predicts the correct classes of the instances in a given test dataset. Robust, \aka adversarial accuracy, estimates how accurately an ML model classifies the generated variants~\cite{Tsipras2019}. 
In this paper, we adopt a point-wise robust accuracy measure, \textit{neighbor accuracy}, to quantify the robustness of a DNN for the neighbors around each data point. 
%In particular, we use two types of spatial transformations (rotation and transformation) % (  rotation as a representative of non-linear transformation and translation as a representative of linear transformation) \todo{is translation linear?} 
%on an image to generate neighbors to measure robust accuracy. 

%Standard accuracy measures how accurately a DNN predicts the correct classes of the instances in a given test dataset. Robust accuracy (also referred as adversarial accuracy) measures how accurate a machine learning model classifies or predicts the class of generated adversarial examples~\cite{Tsipras2019}. In this paper, we adopt a point-wise robust accuracy measure, \textit{neighbor accuracy}, to quantify the robustness of a neural network model for the neighbors (defined in the next section) around each data point. In particular, we use two types of spatial transformations (rotation as a representative of non-linear transformation and translation as a representative of linear transformation) on a data point to generate neighbors and evaluate how accurately a DNN classifies or predicts them to measure the given DNN's robustness at the data point.
%Such settings help us to evaluate the model's robustness in a more realistic setting (as opposed to norm-based perturbation, where an active adversary is needed) and in a controlled fashion (as opposed to GAN).

%% file: theory.tex
%auto-ignore
\section{Methodology}
\label{sec:approach}

% This section describes our approach in detail. Section~\ref{sec:terminology} defines the terminology that we use throughout the paper (\Cref{fig:theory}) followed by the description of our methodology  in~Section \ref{sec:methodology} and \ref{sec:class}.

% \cmt{The above lines seem redundant. The sub-section titles are self-explanatory.}

\subsection{Terminology}
\label{sec:terminology}

\begin{wrapfigure}{r}{0.5\textwidth}
\vspace{-6mm}
\centering
\includegraphics[width=\linewidth]{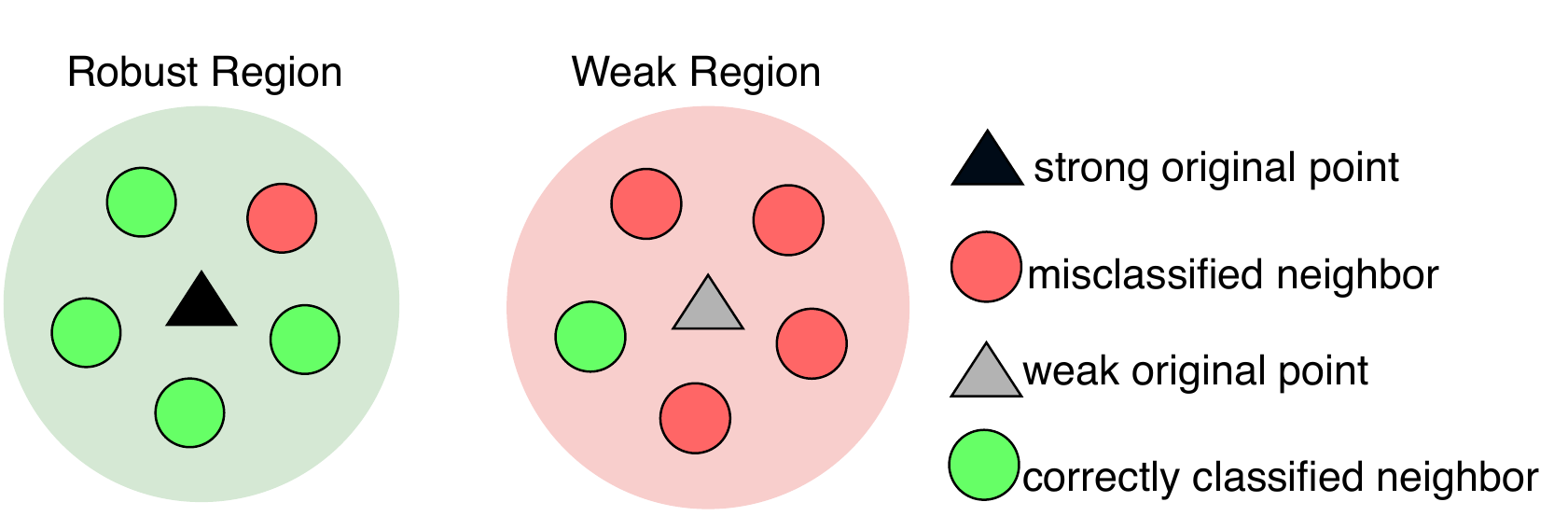}
\caption{\small{\textbf{Illustrating our terminologies. 
The triangles are original points, and the small circles are their neighbors generated by natural variations. 
The light-green region is robust with higher neighbor accuracy, while the light-red region is vulnerable.  %in the left-hand-side, 
The corresponding original points are robust and non-robust  accordingly. 
}}}      
\label{fig:theory}
\vspace{-10mm}
\end{wrapfigure}

%\smallskip
\noindent
\textbf{Original Data Point:} An original data point represents an original un-modified data instance (image in our case) in the studied dataset. The original data points can come from training, validation, or testing dataset, depending on the experimental setting. In Figure~\ref{fig:theory}, the triangle in the center is an original data point. %\cmt{The really long caption of Fig~\ref{fig:theory} is essentially repeating, sentence for sentence the content in this section. As such the figure cannot be understood without reading this section and if one reads it the caption becomes redundant. My recommendation -- delete the caption, except the first sentence, which can be rephrased to be more specific beyond "terminology".}

%An original data point is a data point from a given original dataset. It may come from training set, validation set or testing set. In this paper, the two subjects we study are the object recognition problem and the face recognition problem, so a data point is also an image. We may use data points and images interchangeably. In Figure \ref{fig:theory}, the circle in the center is an original data point.

%\smallskip
\noindent
\textbf{Neighbors:} Neighbors %of an original data point 
are images generated by the natural variations, \eg spatial transformations applied to an original image. Since the transformation parameters are continuous (e.g., degree of rotations), there can be an infinite number of neighbors per image. In Figure~\ref{fig:theory}, the small circles around an original data point represent its neighbors. 
%Neighbors of an original data point are images generated by applying a series of spatial transformations on the original image. There are infinite neighbors for an original image because allowed values for the parameters for spatial transformations are continuous. In Figure \ref{fig:theory}, the five circles around an original data point are neighbors of that original data point. The circles in orange are correctly classified by a pre-trained DNN model under test. The circles in green are mis-classified by the pre-trained DNN model.

%\smallskip
\noindent
%\textbf{Robustness and Neighbor Accuracy:}
\textbf{Neighbor Accuracy:}
%Spatial robustness describes how robustly a DNN model can classify spatially transformed variants of a given image. To measure this, 
We define~\textit{neighbor accuracy} as the percentage of its neighbors, including itself, that can be correctly classified by the DNN model. Figure~\ref{fig:theory} illustrates this; here, red small circles indicate misclassified neighbors, while the green small circles are correctly classified ones. The figure shows that there are only five neighbors per original data point. In the left-hand-side diagram, four out of five neighbors are correctly classified by the given DNN model. If the original data point is correctly classified as well, the neighbor accuracy of the original data is (5/6=) 83.3\%. Similarly, in Figure~\ref{fig:theory} (right), four out of the five neighbors have been misclassified by the model; if the original data point is misclassified, the neighbor accuracy is (1/6=) 16.6\%.

\noindent
%\textbf{Strong (\aka robust)/Weak (\aka non-robust) Points:} 
\textbf{Robustness.}
An original data point is strong, \aka robust, w.r.t.~the DNN model under test if its neighbor accuracy is higher than a pre-defined threshold.
Conversely, a weak, \aka non-robust, point has the neighbor accuracy lower than a pre-defined threshold. For example, at 0.75 neighbor accuracy threshold, the black triangle in~\Cref{fig:theory} is a strong point, and the grey triangle is a weak point.  
%
%\smallskip
%\noindent
%\textbf{Weak/Robust Regions:}

A region contains an original point and all of its neighbors. If the original point is strong (weak), we call the corresponding region as a robust (weak) region. In Figure~\ref{fig:theory}, the light green region is robust while the light red region is weak. 

% A region contains an original point and all of its neighbors. If the original point is strong, we call the corresponding region as a robust region. Conversely, if the original point is weak, the corresponding region is termed as a vulnerable region. As per Figure~\ref{fig:theory}, the light green region in left is robust while the light red region in right is a vulnerable region. 

%As shown in Figure~\ref{fig:robust_region}, all the six points on the left sub-figure are in a robust region while all the six points on the right sub-figure are in a vulnerable region.
%\smallskip
\noindent
\textbf{Neighbor Diversity:} For multi-class classification task,  different neighbors of an original point can be mis-classified to different classes. Neighbor Diversity score measures how many diverse classes a point's neighbors are classified, and is formally computed using Simpson Diversity Index ($\lambda$) \cite{simpson1949}: %, which is defined as
% \begin{equation}
% \label{eq:diversity}
%     \lambda = \sum_{i=1}^{k} p^2_i
% \end{equation}
\inlineequation[eq:diversity]{\lambda = \sum_{i=1}^{k} p^2_i}

where $k$ is the total number of possible classes and $p_i$ is the probability of an image's neighbors being predicted to be class $i$. Large Simpson Index means low diversity. Let's consider we have three possible classes A, B, and C. 
Assume an image has 4 neighbors. Including the original image, there are 5 images in total. If two of the five images are classified as A, and rest are classified as B, then $\lambda=(2/5)^2+(3/5)^2+(0/5)^2=0.52$. In contrast, if two of them are classified as A, and two are classified as B, and one is classified as C then $\lambda=(2/5)^2+(2/5)^2+(1/5)^2=0.36$. Clearly, the latter case is more diverse and thus, has a lower $\lambda$ score.

%\smallskip
\noindent
\textbf{Feature Representation:} 
In a DNN, the neurons' output in each layer capture different abstract representation of the raw input, which are commonly known as features, extracted by the current layer and all the preceding layers. Each layer's output forms the %is in a 
corresponding feature space. 
%In our work, we use the second-to-last layer's feature space, as described in Section~\ref{sec:methodology}, to train our neural network model to predict the robustness of a data point.
For a given input data point, we consider the output  of the DNN's second-to-last layer as its feature representation or feature vector.

%of a data point is the output  of the DNN's second-to-last layer when we feed the data point into the DNN model. 
%We also call the set of the feature representation of all training data representation space.

%We defined two regions, vulnerable regions and robust regions, each of which consists of a neighborhood of points centering an original data point. A vulnerable region is defined to be including a weak point and its neighbors. Similarly, a robust region is defined to be including a strong point and its neighbors. As shown in Figure \ref{fig:robust_region}, all the six points on the left sub-figure are in a vulnerable region while all the six points on the right sub-figure are in a robust region.

%% file: methodology.tex
%auto-ignore
\subsection{Data Collection}
\label{sec:methodology}

% Existing work on spatial robustness focuses on improving the overall robustness of DNNs or analyze them in an aggregated way. 
% In contrast, in this work, we dive deeper into the point-wise robustness property of the model. Here, we describe 
% the data collection and analysis methods we used to answer our research questions.

\noindent
\textbf{Neighbor Generation:} 
For the image classification tasks, for each original image point, we generate its neighbors by combining two types of spatial transformations: rotation and translation. We carefully choose these two types as representatives of non-linear and linear spatial transformations, respectively, following Engstrom et al. ~\cite{engstrom2019exploring}. In particular, following them, we generate a neighbor by randomly rotating the original point by t ($\in [-30, 30]$) degrees, shifting it by $dx$ (about 10\% of the original image's width i.e. $\in [-3, 3]$) pixels horizontally, and shifting it by $dy$ (about 10\% of the original image's height i.e. $\in [-3, 3]$) pixels vertically. It should be noted that for image classification it is standard in the literatures~\cite{rotation2019, engstrom2019exploring,spatial2018} to assume that the transformed image has the same label as the original one.
 As the transformation parameters are continuous, there can be infinite neighbors of an original data point. Hence, we sample $m$ neighbors for each original data point. We explore the impact of $m$ in RQ2.

For the self-driving-car task where the model predicts steering angle, 
for each original image point, we generate 50\% neighbors with rain effect and the rest 50\% with fog effects. We adopt a widely used self-driving car data augmentation package, Automold \cite{automold}, 
for adding these effects where we randomly vary the degrees of the added  effect.
 For the rain effect, we set ``rain\_type=heavy" and everything else as default. For the fog effect, we set everything as default. %\ziyuan{punctuation around heavy does not look right}
%The details of these effects can be found in \cite{automold}.

\noindent
\textbf{Estimating Neighbor Accuracy:} 
%Since a neighbor is a natural variant of an original image, the neighbor should be classified in the same class as given by the labeled data. 
To compute the neighbor accuracy of a data point for a given DNN model, we first generate its neighbor samples by applying different transformations\textemdash spatial for image classification and rain or fog for self-driving-car application. Then we feed these generated neighbors into the DNN model and compute the accuracy by comparing the DNN's output with the label of the original data point. 
For self-driving-car application, we follow the technique described in DeepTest~\cite{tian2017deeptest}. More specifically, if the predicted steering angle of the transformed image is within a threshold to the original image, we consider it as  correct. This ensures that any small variations of steering angle are tolerated in the predicted results.
We then compute $neighbour\ accuracy = \frac{\#correct\  predictions}{original\ point + \#total\ neighbours}$.
%In RQ1 and RQ2, we investigate different spatial robustness properties based on neighbor accuracy. Based on these properties, we further build two methods that can automatically classify strong vs.~weak points and the corresponding region.% (\ie robust vs.~vulnerable region). 
%The following section describes the proposed classification models.

% \begin{figure}[ht]
% \centering
    
% %\includegraphics[width=\linewidth]{experiments/figure/method1aa.pdf}
% \includegraphics[width=0.85\linewidth]{figure/method-Page-2.pdf}
% \caption{\small{\textbf{Classifying Strong vs.~Weak points, and thus, Robust vs.~Vulnerable regions}.
% The three bigger circles on the left side are robust regions. 
% The three bigger circles on the right side are vulnerable regions. The grey dot line is the boundary which our proposed classifiers are trained to learn.}}
% % \cmt{Fonts of the figures are too big compared to the paper text}
% \label{fig:robust_region}
% \vspace{-3mm}
% \end{figure} 

\subsection{Classifying Robust vs.~Weak Points}
\label{sec:class}

We propose two methods, \toolWB\ and \toolBB, to identify whether an unlabeled input is strong or weak \wrt a DNN in real time. 
%the robustness of a DNN model on any unlabeled test image as a strong point or a weak point (in other words, in a robust region or vulnerable region) in real time.
If a test image is identified %by \toolname(either method) 
as a weak point, although it may be classified correctly by the pre-trained model, this image is in a vulnerable region where a slight change to this image may cause the pre-trained DNN to misclassify the changed input.

\subsubsection{\toolWB: White-box Classifier}
\label{sec:toolwb}

\begin{figure}[ht]
\centering
 \vspace{-10mm}
\subfloat[workflow - training\label{fig:workflow_whitebox_train}]{\includegraphics[width=0.48\linewidth]{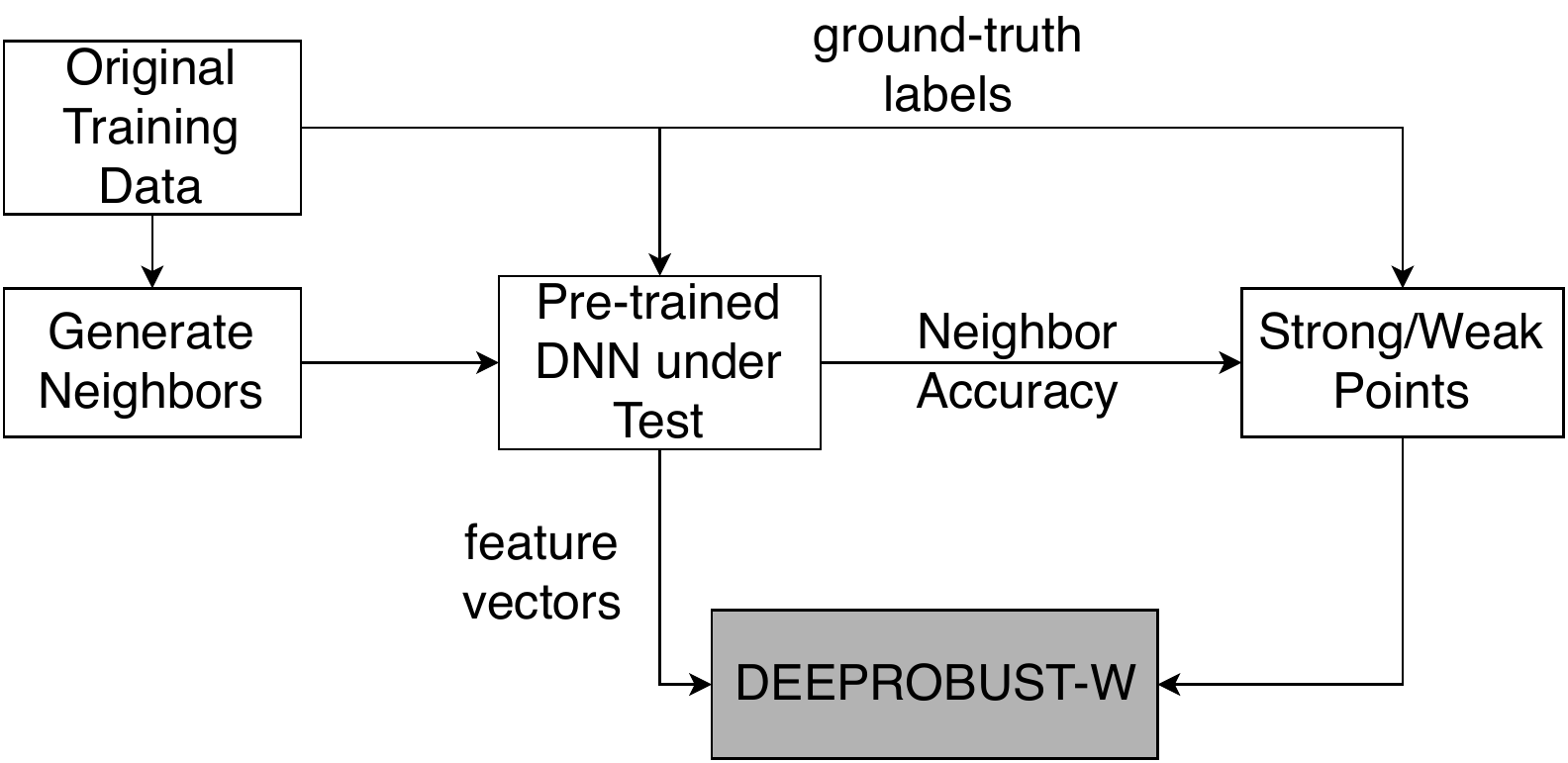}}
\subfloat[workflow - testing\label{fig:workflow_whitebox_test}]{\includegraphics[width=0.48\linewidth]{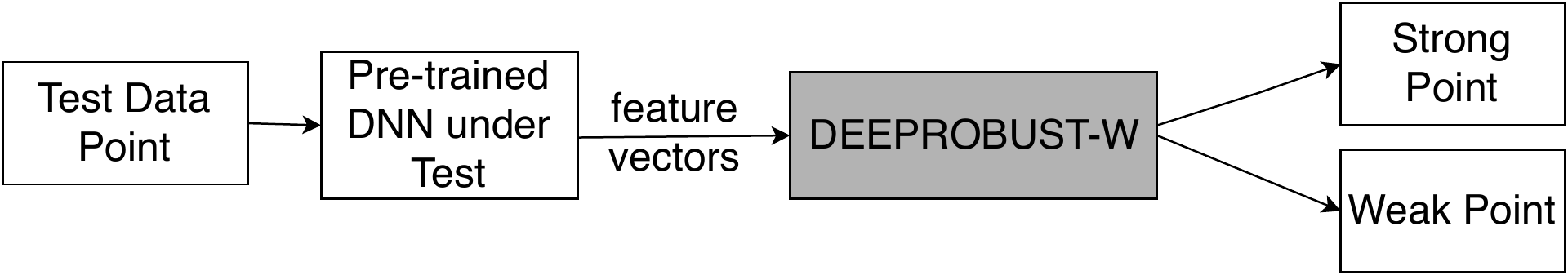}}

\caption{\textbf{\small{Workflow of \toolWB}}}

% \cmt{The font size of Fig. a and b are different.}
\label{fig:workflow_whitebox}  
 \vspace{-7mm}
\end{figure}

This is a binary classifier designed to classify an image (in particular, image feature vector)  as a strong or weak point. Here, we assume that we have white box access to the \mut to extract the feature vectors of the input images from the DNN. 
These feature vectors are given as inputs to \toolWB. Figure~\ref{fig:workflow_whitebox} shows the workflow.

\smallskip
\noindent
\textit{Training}: During training of \toolWB, we first feed all the original training images and their neighbors to the \mut. From the DNN outputs, we compute the neighbor accuracy for each data point in the training set and label each point strong/weak  depending on whether its neighbor accuracy is higher/lower than a predefined threshold. For each original data point, we also extract the output of the DNN's second-to-last layer as its feature vector. 
We use these vectors as inputs to train \toolWB and the outputs are the corresponding  strong/weak labels.

\smallskip
\noindent
\textit{Testing}: Given a test input, we extract its feature vector by feeding the test image to the \mut and then feed the extracted feature vector to the trained \toolWB, which predicts if the input is a strong or weak point.

\begin{wrapfigure}{r}{0.5\textwidth}
\vspace{-15mm}
\centering
\includegraphics[width=\linewidth]{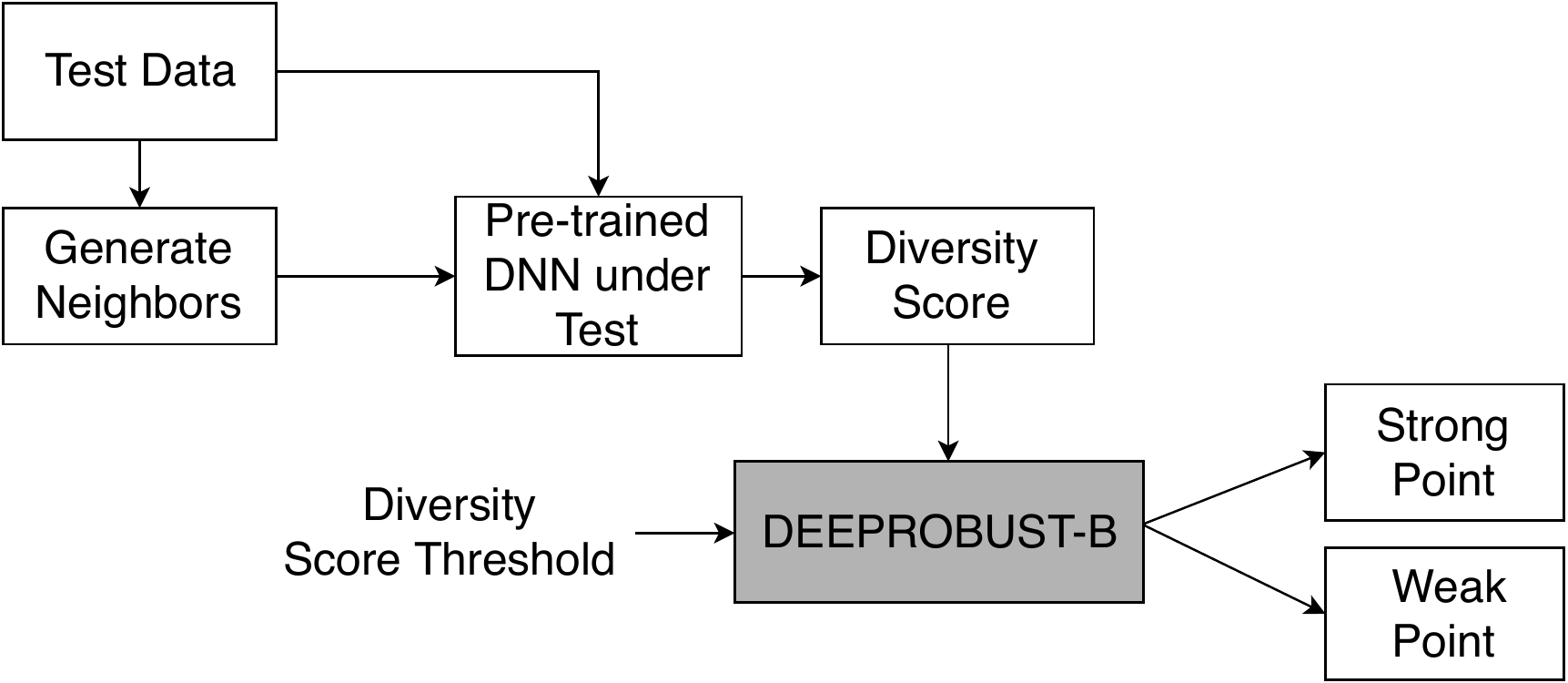}
\caption{\textbf{\small{Workflow of \toolBB}}}
\label{fig:workflow_blackbox}  
\vspace{-10mm}
\end{wrapfigure}

% \begin{figure}[ht]
% \vspace{-10mm}

% \vspace{-15mm}
% \end{figure} 

\subsubsection{\toolBB: Black-box Classifier}
\label{subsec:black_box}

%Similar to \toolWB, \toolBB~
This is also a binary classifier that is intended to classify an image to strong/weak point. 
However, here the user does not have white box access to the \mut. Figure~\ref{fig:workflow_blackbox} shows the workflow.

% \smallskip
% \noindent
% \textit{Training}: No training is needed for this method. 

% \smallskip
% \noindent
% \textit{Testing}: 
Given a test input, we first randomly generate some of its neighbors. We then query the \mut with all these neighbors and compute the diversity score, as per Equation~\ref{eq:diversity}. If the neighbor diversity score (inversely correlated with neighbor diversity) is greater than a given diversity score threshold, the given test input is classified as a strong point; otherwise, a weak point.

Notice that, in this method, we do not need a training step. We only need the diversity score threshold, which can be empirically set using a ground-truth data set. In particular, we first calculate the neighbor accuracy and diversity score of each pre-annotated point. Next, based on a given neighbor accuracy threshold, we identify the weak points, as the ground truth.  The highest diversity score among these weak points is chosen as the diversity score threshold.

% Figure~\ref{fig:workflow_blackbox} shows the workflow of \toolBB. No training procedure is needed for this method. During test time, when given a test input, we first generate some neighbors for it, query the predictions of the pre-trained DNN on them, and then calculate its diversity score. Next, we compare its diversity score with some given diversity score threshold and predict the test input as a strong point if the score is lower than the threshold or in a weak point otherwise.

% One empirical way to set the diversity score threshold is using the training data. In particular, we can first calculate the neighbor accuracy and diversity score of each training point. Next, we sort neighbor accuracy in ascending order and find the percentile of the neighbor accuracy cutoff. Finally, we sort diversity score in ascending order and find the value at the corresponding percentile and adopt this value as the threshold. 

\subsubsection{Usage Scenario} 

\toolname-W/B works in a real-world setting where a customer/user runs a pre-trained DNN model in real-time which constantly receives inputs and wants to test if the prediction of the DNN on a given input can be trusted. 
%\ziyuan{I thought we should say the variation of given input can be trusted.}
\toolWB assumes that the user has white-box access to \mut and all the training data used to train the DNN. 
\toolWB leverages the feature vector and neighbor accuracy of the training data to train the classifier, which can notify the user if the current input is a strong point or weak point. If the input is classified as  strong point, the user can give more trust to the original DNN's prediction. % on the current point and its nearby regions. 
On the other hand, if the point is classified as a weak point, the user may want to be more cautious about the DNN's prediction 
%on the point's surrounding region 
and conduct additional inspections. % of the given input.

In the blackbox setting, \toolBB assumes the user does not have white-box access to \mut. 
\toolBB comes with a small overhead of transforming the input multiple times to get some neighbors and querying \mut on them to estimate the diversity score. 

%% file: experiments.tex
\section{Experimental Design}

%auto-ignore
\subsection{Study Subjects}
\label{sec:experiment}
%\label{sec:exp}
\subsubsection{Image Classification}
%\subsection{Application of DNNs}

%GTSRB\cite{Stallkamp-IJCNN-2011, stallkamp2012man} \\
%GTSRB multi scale model\cite{sermanet2011traffic}\\

%\subsection{Study Subjects}

Similar to many existing works~\cite{tian2017deeptest, zhang2018deeproad, kim2019guiding, ma2018deepgauge, SenETAL05CUTE, tian2019deepinspect} on DNN testing, in this work, we use image classification 
%(\ie object recognition)  
application of DNNs as the basis of our investigation. 
% \cmt{Should we argue that this is also the case with several other works on DNN testing? If so, we'll need to cite those specific papers.} 
This is one of the most popular computer vision tasks, where the model tries to classify the objects in an image or video. %Note that object recognition is different from object detection -- the former requires an image to be classified in different object categories,  while detection also identifies the location of each object. 
% The state-of-the-art DNN models for solving the object recognition problem are variants of convolutional neural networks (CNN)~\cite{he2016deep,wang2017residual,krizhevsky2012imagenet}. % In this work, we explore the properties of our proposed robustness measure neighbor accuracy and evaluate the performance of our robustness classifiers on 
% For our evaluation we use three popular image classification datasets:
% \cifar~\cite{cifar10}, \svhn~\cite{svhn}, and \fmnist~\cite{fmnist}.
% We conduct our experiments on three image classification datasets: \fmnist~\cite{fmnist},  \cifar~\cite{cifar10}, and \svhn~\cite{svhn}. % to answer the research questions described in Section~\ref{sec:res_ques}. 
% We also use popular image classification architectures: \resnet~\cite{he2016deep}, \vgg~\cite{simonyan2014very}, and \wrn~\cite{Zagoruyko2016WRN}.

%\subsection{Study Subject}

\smallskip
\noindent
\textbf{Datasets:} We conduct our experiments on three image classification datasets: \fmnist~\cite{fmnist},  \cifar~\cite{cifar10}, and \svhn~\cite{svhn}.

\begin{itemize}[leftmargin=*]
    \item 
    \textbf{CIFAR-10}: consists of 50,000 training and 10,000 testing 32x32 color images. Each image is one of ten digit classes.

    \item
    \textbf{\fmnist}: consists of 60,000 training images and 10,000 testing 28x28 grayscale images. Each image is one of ten fashion product related classes.

    \item
    \textbf{\svhn}: %This \svhn~(The Street View House Numbers Dataset) is a real world number recognition dataset. \svhn~
    consists of 73,257 training images and 26,032 testing images. Each image is a 32x32 color cropped image of house numbers collected from  Google Street View images. % and belongs to one of the ten digits.

\end{itemize}

\smallskip
\noindent
\textbf{Architectures:} The popular DNN-based image classifiers are variants of convolutional neural networks (CNN)~\cite{he2016deep,wang2017residual,krizhevsky2012imagenet}. 
Here we study the following three architectures for all the three datasets:

\begin{itemize}[leftmargin=*,topsep=0pt]
    \item 
    \textbf{\resnet}: Following Engstrom \etal~\cite{engstrom2019exploring}, we use  \resnet~model with 4 groups of residual layers with filter sizes 16, 16, 32, and 64, and 5 residual units each.
    %We use the same architecture as the one used by Engstrom \etal~\cite{engstrom2019exploring}, which is a \resnet~model with 4 groups of residual layers with filter sizes 16, 16, 32, and 64, and 5 residual units each. 

    \item
    \textbf{\vgg}: We use the same~\vgg~ architecture as proposed in~\cite{simonyan2014very}.  
    
    \item
    \textbf{\wrn}: We use a structure with block type (3, 3) and depth 28 in~\cite{Zagoruyko2016WRN} but replace the widening factor 10 with 2 for less parameters and faster training.
\end{itemize}

\edited{We train all the models from scratch using widely used hyper-parameters and achieve accepted level of validation natural accuracy)}. When training models on \cifar, we pre-process the input images with random augmentation (random translation with $dx, dy \in [-2, 2]$ pixels both horizontally and vertically) which is a widely used preprocessing step for this dataset. When training models on the other two datasets, plain images are directly fed into the models. The natural accuracies and robust accuracies of the models are shown in ~\Cref{tab:models}.

\begin{wraptable}{r}{0.6\columnwidth}
\vspace{-10mm}
\centering

% \footnotesize
\caption{\small{\textbf{Study Subjects (values are in percentage)}}}
\label{tab:models}

\resizebox{0.6\columnwidth}{!}{%
\begin{tabular}{l|l|l|l|l|l|l|l|l|l}
\toprule
Dataset  & \multicolumn{3}{c|}{\cifar} & \multicolumn{3}{c|}{\svhn} & \multicolumn{3}{c}{\fmnist} \\
\midrule
Model  & \vgg & \resnet & \wrn & \vgg & \resnet & \wrn & \vgg & \resnet & \wrn \\

\midrule
nat acc'&  89.0          & 89.3         &    90.6       &      94.5     &     95.3      &    95.2  &  93.4      &   93.5        & 93.6     \\
rob acc* &  75.5          & 68.5         &     74.8      &     78.1      &   78.9        &    81.  &  61.1      &   63.0        & 64.2     \\ 
\bottomrule
\end{tabular}
}
{\scriptsize\\'Natural accuracy. *Robust accuracy is estimated as the average neighbor accuracy for test data points.}
\vspace{-10mm}
\end{wraptable}

\subsubsection{Steering Angle Prediction}
  
We further evaluate \toolWB in a self-driving car application to show that it can be applied into a regression task. % and an application besides image classification. 
These models learn to steer (\ie predict steering angle) by taking in visual inputs from car-mounted cameras that record the driving scene, paired with the steering angles from a human driver. 

% \todo{Ziyuan: should we mention behavior cloning here. I feel it is more confusing.}

\smallskip
\noindent
\textbf{Datasets:}
{We use the dataset by Stocco \etal  \cite{2020-icse-misbehaviour-prediction}, which is collected by the  authors driving on three tracks of different environments in the Udacity Simulator \cite{udacity-simulator}. It consists of 37888 central camera training images and 9427 central camera evaluation images. Each image is of size 320x120.}

\smallskip
\noindent
\textbf{Architectures:}
{We evaluate our method on the three pre-trained DNN models used in \cite{2020-icse-misbehaviour-prediction}: NVIDIA DAVE-2 \cite{nvidia-dave2}, Epoch \cite{epoch}, and Chauffeur \cite{chauffeur}. These models have been used by many previous testing works on self-driving car \cite{pei2017deepxplore, tian2017deeptest, 2020-icse-misbehaviour-prediction}.} 

\subsection{Evaluation}
\label{subsec:eval}

\noindent
\textbf{Evaluation Metric.} {We evaluate both \toolWB and \toolBB   for detecting weak points under twelve and nine different DNN-dataset combinations, respectively, in terms of precision, recall, and F1 score. Let us assume that $E$ is the number of weak points detected by our tool and $A$ is the the number of true weak points in the ground truth set. Then the precision and recall are $\frac{|A\cap E|}{|E|}$ and $\frac{|A\cap E|}{|A|}$, respectively. F1 score is a single accuracy measure that considers both precision and recall, and defined as $\frac{2\times precision\times recall}{precision+recall}$. We perform each experiment for two thresholds of neighbor accuracy that defines strong vs. weak points: 0.75 and 0.50.} 

\noindent
\textbf{Baselines.}
We compare \toolWB and \toolBB with two baselines. 
One naive baseline (denoted {\em random}) is randomly selecting the same 
number of points as detected by our proposed method to be weak points. 
Another baseline (denoted {\em top1}) is based on prediction confidence score\textemdash if the confidence of a data point is higher than a pre-defined cutoff we call it a strong point, weak otherwise. 
%predicting those data points with confidence score of the predicted class higher than the cutoff to be strong and others weak. 
This baseline is based on the intuition that DNNs might not be confident enough to predict the weak points.

%% file: rq2.tex
\section{Results}
\label{sec:result}

In this section, we elaborate on our results.  In our preliminary experiments, we have two findings regarding neighbor accuracy. First, the neighbor accuracy vary widely across data points and there is a non-trivial number of points having relatively low neighbor accuracy. For example, for all the models trained on \cifar~dataset, 40\% of training data and 42\% of testing data have neighbor accuracy <0.75, and 16\% of training data and 20\% of testing data have neighbor accuracy <0.50. These points degrade the aggregated spatial robustness of the model. The same finding holds for the other two datasets. Second, the distribution of neighbor accuracy for a dataset is similar across different models. For \cifar, \fmnist~and \svhn, $60\%$, $76\%$, and $81\%$, respectively, of data points have neighbor accuracy change $<0.2$ across any two models on the same dataset. This implies that a large portion of data points' neighbor accuracy is independent of the model selected. Detailed results can be found in \Cref{appendix}. 

The first observation shows that neighbor accuracy is a distinguishable measure for local robustness for the datasets and models we study. The second observation implies that the properties of points of low neighbor accuracy may be similar across models for each dataset. Following these two observations, we dive deeper and explore the characteristics of data points with different neighbor accuracy in RQ1. We then evaluate the performance of \toolWB and \toolBB which are developed based on the observations from RQ1 in RQ2 and RQ3, respectively. Finally, in RQ4, we evaluate the generalizability of our method by applying \toolWB in a regression task for self-driving cars under more complex transformations.

% The first three RQs are based on the controlled study, where we spatially transformed the images with varying degrees of spatial transformations for analyzing image classification models. 
% RQ1 presents our empirical findings on the different characteristics of local robustness. 
% In RQ2 and RQ3, we evaluate \toolWB and \toolBB in identifying the weak points.
% In RQ4, we evaluate the generalizability of our method by applying \toolWB in a regression task for self-driving cars under more complex transformations.

\smallskip

\RQ{1}{\rqb}%{What are the characteristics of non-robust data points in their feature space?}
%What are the distribution/characteristics of weak points?}
\label{sec:rq2}

We explore the characteristics of robust vs.~non-robust points in their feature space. In particular, we check the difference in feature representations between:
a) robust and non-robust points, and b) points with different degrees of robustness.
% in the feature space whether a non-robust point is different from: 

% \renewcommand{\theenumi}{\alph{enumi}}
% %\begin{itemize}
% \begin{enumerate}[label=\alph*)] 
%     \item robust and non-robust points;
%     \item points with different degrees of robustness.
%     % \item whether there are any differences between robust and non-robust points in the feature space;
%     % \item whether there are differences between points with different degrees of robustness.
% \end{enumerate}

% ~\ziyuan{Define feature space}
% I have defined it in Sec 4.1
%of spatial robustness across different data points of the studied data sets w.r.t. three well-trained models. We measure spatial robustness in terms of {\em neighbor accuracy}. In particular, we explore the variations of spatial robustness:
\smallskip
\RQA{1a}{Given a well trained model, do the feature representations of robust and non-robust points vary?} In this RQ, we first explore how robust (\ie strong) and non-robust (\ie weak) data points are distributed in the feature space. % such that we might train a classifier to differentiate them from strong points.

\begin{figure}
\vspace{-5mm}
\centering
\subfloat[\cifar]{\includegraphics[width=0.22\columnwidth]{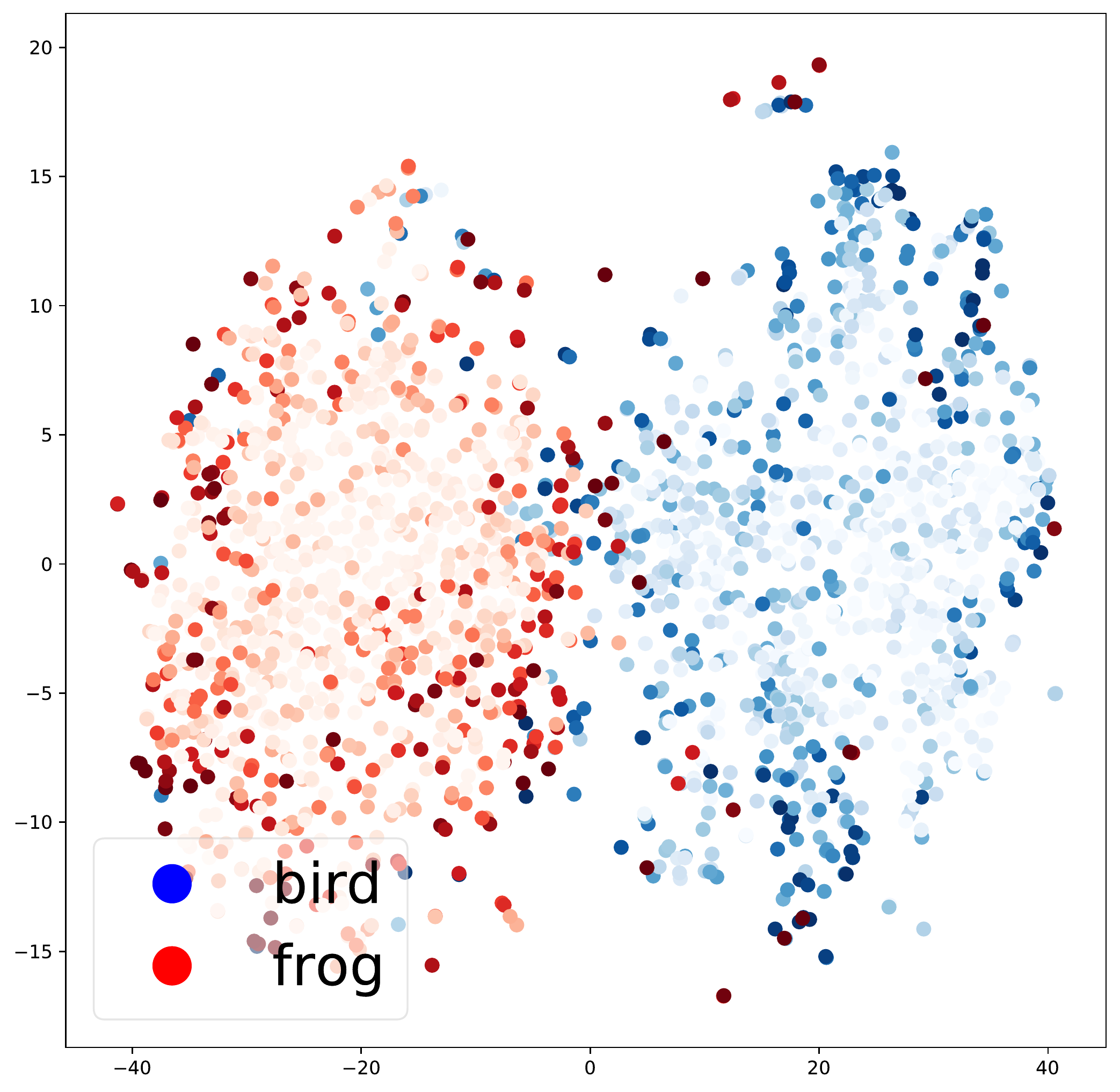}}
~
\subfloat[\fmnist]{\includegraphics[width=0.22\columnwidth]{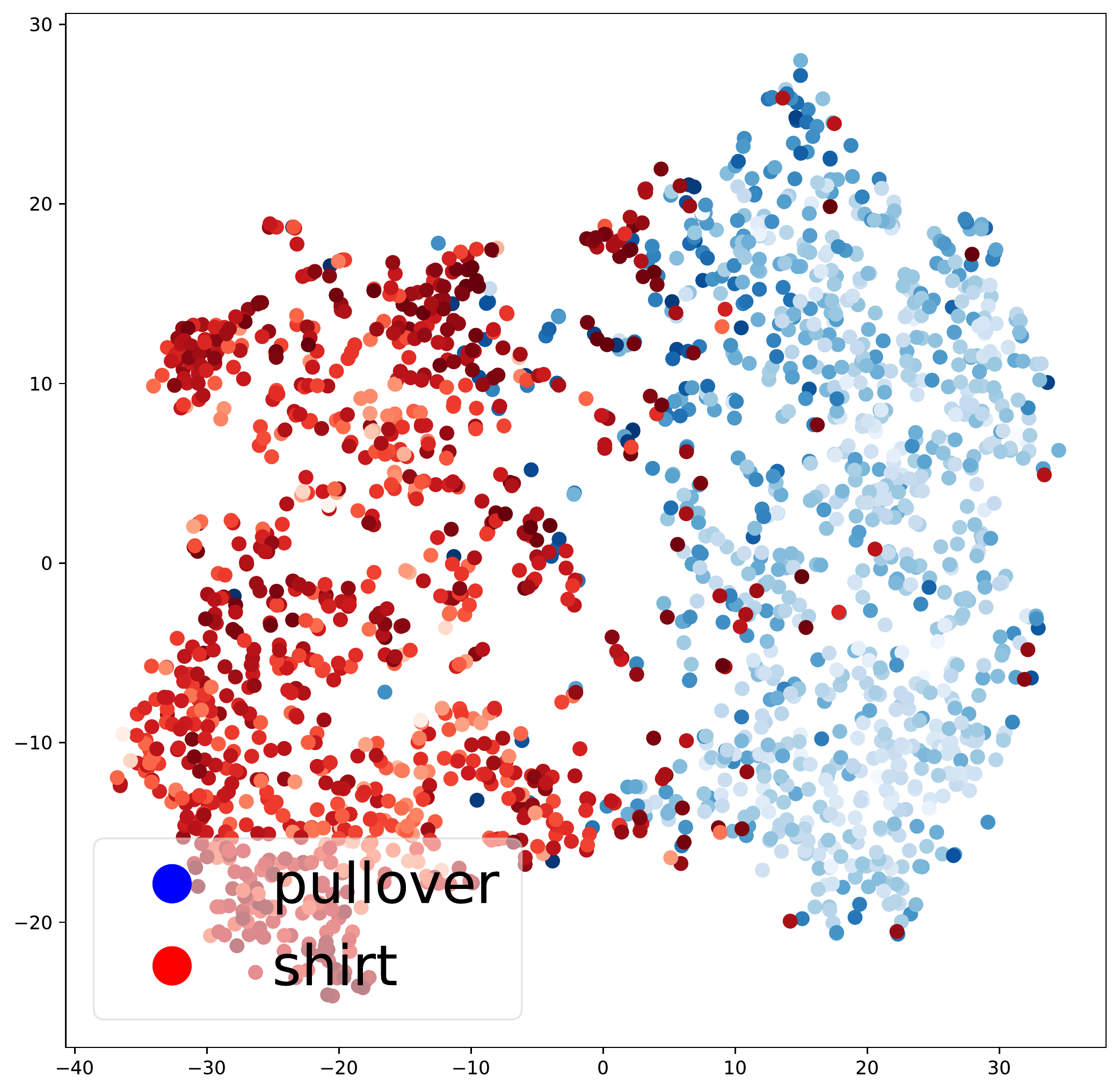}}
~
\subfloat[\svhn]{\includegraphics[width=0.22\columnwidth]{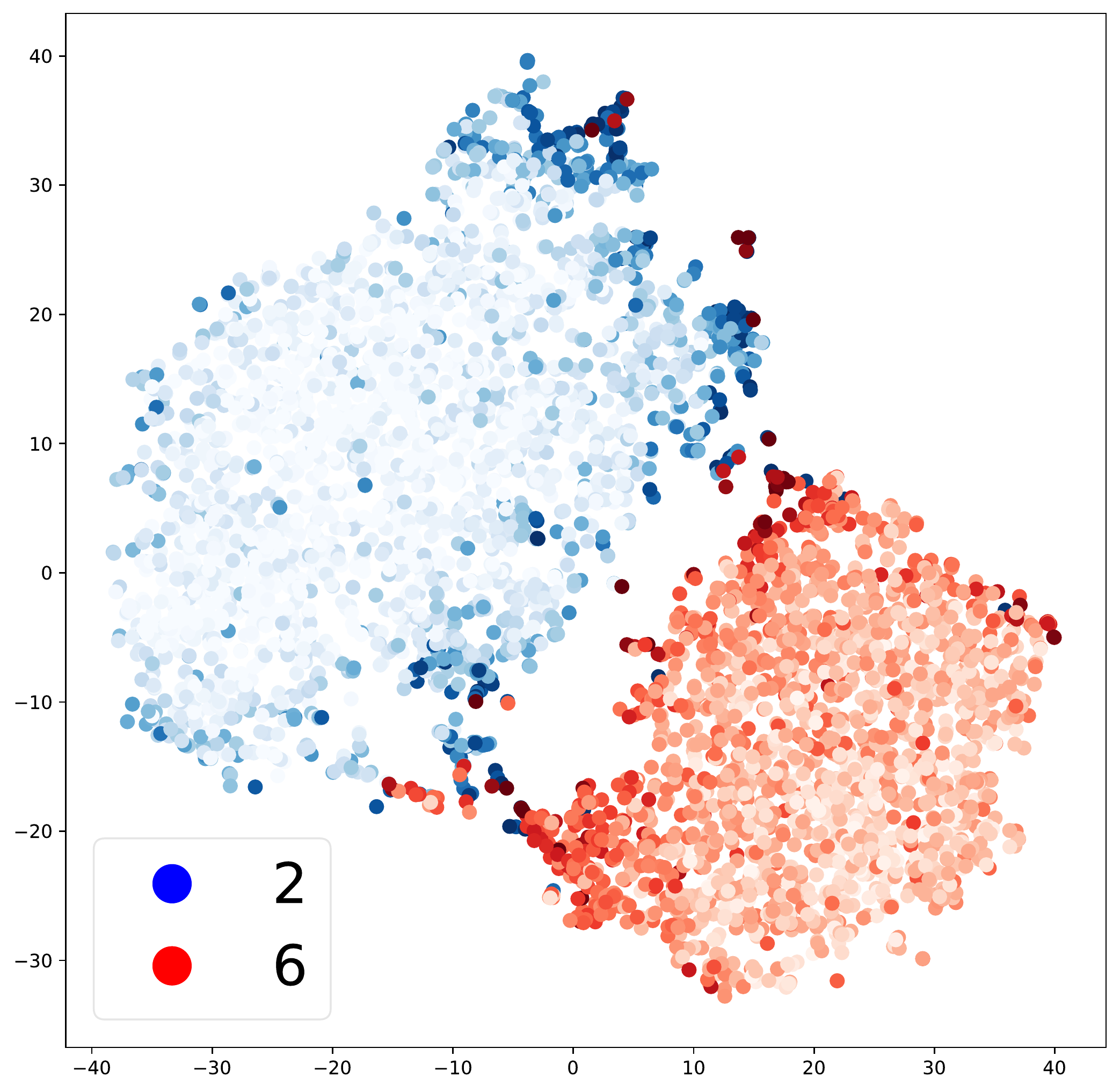}}
\caption{\textbf{\small{The t-SNE plots of data points from two randomly chosen classes across three datasets using ResNet. 
%The t-SNE plots of data points from two randomly chosen classes across all three datasets. 
Darker color indicates lower neighbor accuracy. 
%The darker  the color of a point, the lower the neighbor accuracy the point has
}}}
\label{fig:tsne_two_classes}
\vspace{-8mm}
\end{figure}

We apply t-SNE\cite{vanDerMaaten2008}, a widely used visualization method, to visualize the distribution of points of different neighbor accuracy in the representation space for all three datasets when using \resnet~as the classifier. Figure \ref{fig:tsne_two_classes} shows the visualization of feature vectors from two randomly picked classes with colors indicating the neighbor accuracy of each point. The darker a point's color is, the lower its neighbor accuracy is. It is evident that most points of low neighbor accuracy tend to be further away from the class center. 

% Figure~\ref{fig:tsne_all_classes} further shows the visualization of feature vectors of all ten classes. Each class has two colors indicating weak/strong points. We set a threshold of neighbor accuracy at 0.5 to mark the strong points. Although the visualization is slightly noisier than that for two classes, the general pattern that most weak points are further away from the corresponding class center still holds.

To numerically verify this observation, first, we define a class center $c_k$ for each class $k$ as the median value of the feature vectors of all the points from class $k$. Thus, if $f_{i}$ is the feature of a point at i$^{th}$ dimension and $\median{f_{ik}}$ is the median of the i$^{th}$ dimension features for all the  points in class $k$, $c_k$ is defined to be $(\median{f_{1k}}, ..., \median{f_{jk}}, ..., \median{f_{nk}})$.

% {
% \setlength{\abovedisplayskip}{0pt}
% \setlength{\belowdisplayskip}{0pt}
% \vspace{-0.3cm}
% \begin{equation}
% \scriptstyle
% \label{eq:napm}
%   c_k=
% \begin{blockarray}{c}
% %  & C_1 & ... & C_i & ... & C_m \\
% \begin{block}{(c)}
%   \median{f_{1k}}  \\
%   ...   \\
%   \median{f_{jk}}  \\
%   ...  \\
%   \median{f_{nk}} \\
% \end{block}
% \end{blockarray}
% \end{equation}
% \vspace{-0.3cm}
% }

The reason we take median rather than mean is that it is a more statistically stable measure and is less likely to be heavily influenced by outliers in the representation space. Then, for every point $p$, %(e.g. with index $i$), 
we define a ratio: $    %r^{(i)} = \frac{d^{(i)}_1}{d^{(i)}_2}
    r^{(p)} = \frac{d^{(p)}_{same\_class}}{d^{(p)}_{nearest\_other\_class}}$, 
where {$d^{(p)}_{same\_class}$ is the distance of the $p$-th point's feature vector to its own class center} and $d^{(p)}_{nearest\_other\_class}$ is the distance of the $p$-th point's feature vector to the class center of its closest other class. A small $r^{(p)}$ means that the point $p$ is close to its own class center while far from other classes, \ie $p$ is far from the decision boundary. In contrast, a larger $r^{(p)}$ indicates that the point $p$ is closer to some other classes, \ie it is closer to the decision boundary.

% While measuring the above ratio, a class center $c_k$ for a class $k$ is defined as the median value of the feature vectors of all the points from class $k$ at every dimension. The reason we take median rather than mean is that it is a more statistically stable measure and is less likely to be heavily influenced by outliers in the representation space. 
% ~\ziyuan{It'd be probably better to show the boxplot than average---boxplot can show the entire range.} 

\begin{wraptable}{r}{0.6\columnwidth}
\vspace{-12mm}
\centering
\footnotesize
 \caption{\small{\textbf{Weak and strong points ratio, and cohen's d effect size}}}
\label{tab:effect_size_5}

\resizebox{0.6\columnwidth}{!}{%
\begin{tabular}{l|l|l|l|l|l|l|l|l|l}
\toprule
Dataset  & \multicolumn{3}{c|}{\cifar} & \multicolumn{3}{c|}{\svhn} & \multicolumn{3}{c}{\fmnist} \\
\midrule
Model  & \resnet & \wrn & \vgg & \resnet & \wrn & \vgg & \resnet & \wrn & \vgg \\
\midrule
\multicolumn{10}{c}{Neighbor Accuracy Cutoff=0.5} \\
\midrule
$r_w$ & 0.915 & 0.955 & 1.004 & 1.046 & 1.103 & 0.997 & 0.746 & 0.734 & 0.976 \\
$r_s$ & 0.609 & 0.584 & 0.975 & 0.294 & 0.309 & 0.977 & 0.297 & 0.293 & 0.930 \\
d* & 1.368 & 1.736 & 1.163 & 2.077 & 2.428 & 1.420 & 1.426 & 1.312 & 1.332 \\
\bottomrule

%\toprule
\multicolumn{10}{c}{Neighbor Accuracy Cutoff=0.75} \\
%Dataset  & \multicolumn{3}{c|}{\cifar} & \multicolumn{3}{c|}{\svhn} & \multicolumn{3}{c}{\fmnist} \\
% \midrule
% Model  & \vgg & \resnet & \wrn & \vgg & \resnet & \wrn & \vgg & \resnet & \wrn \\

\midrule
$r_w$ & 0.778 & 0.796 & 0.992 & 0.604 & 0.671 & 0.983 & 0.516 & 0.496 & 0.953 \\
$r_s$ & 0.588 & 0.558 & 0.973 & 0.260 & 0.274 & 0.977 & 0.253 & 0.257 & 0.918 \\
d* & 0.786 & 1.040 & 0.749 & 0.860 & 1.111 & 0.401 & 0.749 & 0.642 & 0.937 \\
\bottomrule

\end{tabular}
}

{\scriptsize *Cohen's d effect size of 0.20 = small, 0.50 = medium, 0.80 = large, 1.20 = very large, and 2.0 = huge~\cite{cohen1988spa,Sawilowsky09}.}
\vspace{-10mm}
\end{wraptable}

We then measure the average $r^{(p)}$ among the weak points (denoted as $r_w$) 
and among strong points (denoted as $r_s$) 
for all three datasets across three models. Besides, we also calculate mann-whitney wilocox test\cite{Mann47} and cohen's d effect size \cite{cohen1988spa} between the two ratios to test if the two ratios indeed have statistically significant difference and how large the difference is.

%$W$ is the set consists of all the indices for weak points, $S$ is the set consists of all the indices for strong points,
% \begin{table}[htpb]
% \vspace{-6mm}

% \end{table}

% \cmt{incomplete}
%According to \cite{cohen1988spa,Sawilowsky09}, for Cohen's d effect size, 0.20 means small, 0.50 means medium, 0.80 means large, 1.20 means very large, and 2.0 means huge. 
As shown in Table \ref{tab:effect_size_5}, for both the neighbor accuracy cutoff (0.5 and 0.75), except one setting, the cohen's d effect size for every setting is larger than 0.50, which implies a medium to very large difference. Besides, for every setting, the mann-whitney wilocox test value (not shown in the table) is smaller than $1e^{-80}$, which implies the difference is indeed statistically significant.

The visualization and numerical results imply that most weak points are close to the decision boundaries between classes. % and thus can be differentiated from strong points. 
Note that similar observation was also observed by Kim et. al.~\cite{kim2019guiding} in case of adversarial perturbation. 
In particular, they find that adversarial examples tend to be closer to class decision boundaries. In contrast, we focus on spatial robustness and find that spatially non-robust points are closer to decision boundaries.
% ~\ziyuan{similar observation has been observed in surprise adequacy paper for adversarial perturbation}

%\RS{2a}{ In the representation space, weak points tend to lie towards the class decision boundary while the strong points lie towards the center.} 

\smallskip

\RQA{1b}{Given a well trained model, do the feature representations of the data points vary by their degree of robustness?} 
%Since weak points are more likely to be closer to a decision boundary, a natural follow-up question to ask is that if the given classifier shows different behaviors on the surrounding regions of weak points (a.k.a. vulnerable regions) and those of strong points (a.k.a. robust regions).
By analyzing the classifications of the neighbors of weak vs.~strong points, %we performance of the given classifier on the weak points and strong points respectively, 
we observe that the weaker a point is, its neighbors are more likely to be classified in different classes.  
%compared with strong points, weak points tend to be confused by a classifier to be more diverse classes. 
We quantify this observation by computing diversity of the outputs a point's neighbor; We adopt Simpson Diversity Index ($\lambda$) \cite{simpson1949} as defined in~\Cref{eq:diversity}.
% which is defined as $\lambda = \sum_{i=1}^{k} p^2_i$, 
% where $k$ is the total number of possible classes and $p_i$ is the probability of an image's neighbors being predicted to be class $i$. Large Simpson index means low diversity. Let's consider we have three possible classes A, B, and C. 
% Assume an image has 4 neighbors. Including the original image, there are 5 images in total. If two of the five images are classified as A, and rest are classified as B, then $\lambda=(2/5)^2+(3/5)^2+(0/5)^2=0.52$. In contrast, if two of them are classified as A, and two are classified as B, and one is classified as C then $\lambda=(2/5)^2+(2/5)^2+(1/5)^2=0.36$. Clearly, the latter case is more diverse and thus, low $\lambda$ score.

% \begin{table}[!htpb]
\begin{wraptable}{r}{0.6\columnwidth}
\vspace{-8mm}
\setlength{\tabcolsep}{2pt}
\centering
\footnotesize
\caption{\small{\textbf{Spearman Correlation between Neighbor Accuracy and Simpson Diversity Index. All coefficients are reported with statistical significance ($p<0.05$).}}}
\label{tab:diversity_cor}
%\begin{tabular}{l|l|l|l|l|l|l|l|l|l}

\resizebox{0.6\columnwidth}{!}{%
\begin{tabular}{llll|lll|lll}
\toprule
Dataset  & \multicolumn{3}{c|}{\cifar} & \multicolumn{3}{c|}{\svhn} & \multicolumn{3}{c}{\fmnist} \\
\midrule
Model   & \resnet & \wrn & \vgg & \resnet & \wrn & \vgg & \resnet & \wrn & \vgg \\

\midrule
corr.coeff. &  0.853          & 0.909         &    0.946       &      0.970     &     0.984      &    0.983       &    0.923      &   0.962       &  0.8947 \\
\bottomrule 

\vspace{-14mm}
\end{tabular}
}
\end{wraptable}
% \end{table}

Table~\ref{tab:diversity_cor} shows the Spearman correlation between neighbor accuracy and $\lambda$ on the three datasets and three models for each. Note that while calculating the correlation, we remove points with neighbor accuracy 100\% since there are many points having 100\% neighbor accuracy and tend to bias upward the Spearman Correlation; if we include points with neighbor accuracy 100\%, the correlations become even higher.
We notice that for any setting, the Spearman Correlation is never lower than 0.853. This indicates that neighbor accuracy and diversity are highly correlated with each other. For example, the bird image in Fig.\ref{fig:bird1} has neighbor accuracy 0.49 and diversity 0.36, while the bird image in Fig.\ref{fig:bird2} has neighbor accuracy 1 and diversity 1.
This shows, the classifier tends to be confused about weak points and mispredicts them into many different kinds of classes.

% \RS{2}{Weak points tend to have a larger distance from its class's center in the representation space. An image with lower neighbor accuracy tend to be confused to be more diverse classes. }
%\RS{2b}{The weaker an image is, the model under test tends to be more confused by it, and classify its neighbors into more diverse classes.}

\smallskip
\RS{1}{In the representation space, weak points tend to lie towards the class decision boundary while the strong points lie towards the center. The weaker an image is, the model tends to be more confused by it, and classify its neighbors into more diverse classes.}
\smallskip

%% file: rq3.tex
%auto-ignore
%\smallskip
\RQ{2}{\rqc}
\label{sec:rq3}

We explore this RQ using \toolWB, as discussed in~\Cref{sec:toolwb}. \toolWB takes the feature vector of a data point as input and classifies it to a strong/weak point. We implement \toolWB with a simple 4-layer, fully connected neural network architecture with hidden layer dimensions 1500, 1000, and 500, respectively.

Table \ref{tab:diff_architecture} shows the result. At $0.75$ setting, \toolWB has F1 up to $91.4\%$, with an average of $76.9\%$.  At $0.50$ setting, \toolWB detects weak points with average F1 of $61.1\%$, while it can go up to $79.1\%$. \toolWB consistently performs significantly better than the baseline methods.

The top1 has very good precision, since a mis-classified image with low confidence tends to have very poor local robustness. However, there also exist many images that are correctly classified with high confidence yet have poor local robustness. The miss of these points leads the top1 to have very poor recall and thus even worse F1 compared with the random baseline. Our method comes to aid by providing high recall at the same time of decent precision.

\begin{wraptable}{l}{0.45\columnwidth}
\vspace{-6mm}
\centering
      \setlength{\tabcolsep}{3pt}
      \caption{\small{\textbf{Performance of \toolWB and the baseline methods for predicting weak points.}}}
      \label{tab:diff_architecture}
        
    \resizebox{0.45\columnwidth}{!}{%
        \begin{tabular}{l|l|l|rrr|rrr}
        \toprule
        
         dataset  & model & method & \multicolumn{3}{c|}{0.75 neighbor acc.} & \multicolumn{3}{c}{0.50 neighbor acc.} \\
        \midrule
        
          &   &  & f1 & tp & fp & f1 & tp & fp \\
            \midrule
        \cifar & \resnet & ours & 0.79 & 3844 & 764 & 0.581 & 1290 & 664
     \\
             & & top1 & 0.376 & 1218 & 206 & 0.182 & 255 & 120 \\
             & & random & 0.488 & 2372 & 2236 & 0.233 & 520 & 1445 \\
                \cmidrule{2-9}
             & \wrn & ours & 0.747 & 2901 & 906 & 0.56 & 947 & 610 \\
             & & top1 & 0.35 & 889 & 222 & 0.183 & 189 & 90 \\
             & & random & 0.395 & 1534 & 2273 & 0.154 & 261 & 1296 \\
                 \cmidrule{2-9}
             & \vgg & ours & 0.654 & 2222 & 938 & 0.493 & 747 & 543 \\
             & & top1 & 0.439 & 1070 & 153 & 0.266 & 278 & 106 \\
             & & random & 0.332 & 1127 & 2033 & 0.132 & 200 & 1090 \\
            \midrule
        \svhn & \resnet & ours & 0.755 & 6814 & 2530 & 0.577 & 1414 & 674 \\
             & & top1 & 0.315 & 1665 & 142 & 0.267 & 452 & 122 \\
             & & random & 0.343 & 3095 & 6249 & 0.086 & 210 & 1878 \\
                 \cmidrule{2-9}
             & \wrn & ours & 0.709 & 5062 & 2143 & 0.582 & 1404 & 1055 \\
             & & top1 & 0.292 & 1238 & 130 & 0.203 & 275 & 85 \\
             & & random & 0.28 & 2000 & 5205 & 0.095 & 229 & 2230 \\
                 \cmidrule{2-9}
             & \vgg & ours & 0.595 & 5214 & 3367 & 0.498 & 1272 & 911 \\
             & & top1 & 0.172 & 840 & 67 & 0.139 & 221 & 52 \\
             & & random & 0.341 & 2986 & 5595 & 0.094 & 240 & 1943 \\
            \midrule
        \fmnist & \resnet & ours & 0.914 & 6034 & 873 & 0.791 & 2144 & 556 \\
             & & top1 & 0.124 & 428 & 11 & 0.039 & 57 & 7 \\
             & & random & 0.657 & 4340 & 2567 & 0.263 & 712 & 1988 \\
                 \cmidrule{2-9}
             & \wrn & ours & 0.896 & 5743 & 652 & 0.76 & 2033 & 641 \\
             & & top1 & 0.144 & 490 & 14 & 0.045 & 63 & 8 \\
             & & random & 0.638 & 4093 & 2302 & 0.281 & 752 & 1922 \\
                 \cmidrule{2-9}
             & \vgg & ours & 0.864 & 6348 & 1231 & 0.654 & 1895 & 1082 \\
             & & top1 & 0.104 & 392 & 5 & 0.028 & 39 & 5 \\
             & & random & 0.734 & 5393 & 2186 & 0.295 & 854 & 2123 \\
      \bottomrule
        \end{tabular}
        }
\vspace{-10mm}
\end{wraptable}

Notice that \toolWB's performance depends on the training data selection, mainly (a) how many weak vs.~strong points are used to train the model, and (b) how many neighbors are generated per point to decide whether it is strong/weak.  To investigate the previous one, we assign a weight to each input point, indicating how likely it would be selected to train \toolWB. In particular, for an input $i$, a weight $w_i := \frac{1+(1-n_i)^m \times 100^m}{1+100^m}$ is computed, where $n$ is its neighbor accuracy, and $m$ is a configurable parameter; with larger $m$, more weak points are sampled, and vice versa. Thus, if $m$ is larger, \toolWB will be trained with more weak points and vice versa.

Table~\ref{tab:diff_exp}A shows the performance: as $m$ increases, the detector trades precision for recall. In this way, choosing different values of $m$, the precision-recall trade-off of the detector can be adjusted according to a user's need. {From a different perspective, this way of oversampling weak points also addresses the potential problem of imbalanced data when the weak points are much less than the strong points.}
\input{table_rq3_2.tex}

Next, we check how \toolWB's performance is dependent on the number of sampled neighbors, because a data point can potentially have infinite neighbors. Table~\ref{tab:diff_exp}B shows that the number of neighbors does not have much influence on the performance of the detector once it goes beyond some value (F1 score change less than 3.5 percentage point between 25 and 200 samples) for all the three datasets. Thus, we choose 50 for all of our experiments. \edited{For future work, a statistical bound with confidence intervals for neighbor accuracy can be estimated by modeling neighbor accuracy using distributions like folded normal.}

\smallskip
\RS{2}{\toolWB can identify weak points with reasonably high F1 score: on average $76.9\%$, at $0.75$ neighbor accuracy cut-off.}

%% file: table_rq3_2.tex
%auto-ignore
 \begin{table}[ht]
%\begin{wraptable}{r}{0.95\columnwidth}
% \vspace{-10mm}
  \vspace{-12mm}
\setlength{\tabcolsep}{2pt}
\centering
% \scriptsize
 \caption{\small{\textbf{\toolWB performance using different sampling strategies for training}}}
\label{tab:diff_exp}%
\resizebox{0.8\columnwidth}{!}{%

  \begin{tabular}{l|r|rrrrr}
    \toprule
    \multicolumn{7}{c}{\small{\bf A: with varying number of strong/weak points}} \\
    \toprule
    dataset & m & prec & recall & tp & fp & f1\\
        \midrule
    \cifar 
         & 0 & \textbf{0.660} & 0.518  & 1290 & 664 & 0.581\\
         & 1 & 0.615 & 0.599  & 1490 & 932 & 0.607\\
         & 2 & 0.544 & \textbf{0.699}  & 1740 & 1460 & \textbf{0.612}\\
        \midrule
    \svhn 
        & 0 & \textbf{0.677} & 0.502  & 1414 & 674 & 0.577\\
        & 1 & 0.575 & 0.653  & 1837 & 1357 & \textbf{0.612}\\
        & 2 & 0.332 & \textbf{0.767}  & 2160 & 4356 & 0.463\\
        \midrule
    \fmnist 
        & 0 & \textbf{0.794} & 0.787  & 2144 & 556 & \textbf{0.791}\\
        & 1 & 0.746 & 0.839  & 2284 & 777 & 0.79\\
        & 2 & 0.712 & \textbf{0.871}  & 2372 & 962 & 0.783\\
  \bottomrule
    \end{tabular}%
% \largeskip~~~~~~~~~
\noindent
  \begin{tabular}{l|r|rrrrr}
    \toprule
    \multicolumn{7}{c}{\small{\bf B: with varying number of neigbours}} \\
    \toprule
    dataset & \#neighbors & prec & recall & tp & fp & f1\\
        \midrule
    \cifar 
         & 6 & 0.662 & 0.389 & 967 & 493 & 0.49 \\
         & 12 & 0.685 & 0.384 & 955 & 440 & 0.492 \\
         & 25 & 0.665 & 0.502 & 1250 & 629 & 0.572 \\
         & 50 & 0.660 & 0.518 & 1290 & 664 & 0.581 \\
         & 200 & 0.683 & 0.507 & 1261 & 585 & 0.582 \\
        \midrule
    \svhn 
        & 6 & 0.723 & 0.403 & 1136 & 436 & 0.518 \\
        & 12 & 0.672 & 0.527 & 1483 & 725 & 0.59 \\
        & 25 & 0.619 & 0.629 & 1771 & 1090 & 0.624 \\
        & 50 & 0.632 & 0.605 & 1703 & 993 & 0.618 \\
        & 200 & 0.667 & 0.550 & 1550 & 774 & 0.603 \\
        \midrule
    \fmnist 
        & 6 & 0.817 & 0.727 & 1981 & 443 & 0.77 \\
        & 12 & 0.784 & 0.790 & 2153 & 592 & 0.787 \\
        & 25 & 0.773 & 0.787 & 2143 & 629 & 0.78 \\
        & 50 & 0.836 & 0.727 & 1981 & 390 & 0.778 \\
        & 200 & 0.778 & 0.812 & 2211 & 632 & 0.794 \\
  \bottomrule
    \end{tabular}%

}
    \vspace{-8mm}
%\end{wraptable}
 \end{table}

% \begin{tabularx}{0.95\linewidth}{l|lll|lll|lll}
% \toprule
% \multicolumn{10}{c}{\small{\bf A: with varying number of strong/weak points}} \\
% \toprule
% Dataset  & \multicolumn{3}{c|}{\cifar} & \multicolumn{3}{c|}{\svhn} & \multicolumn{3}{c}{\fmnist} \\
% \midrule
% m  & 0 & 1 & 2 & 0 & 1 & 2 & 0 & 1 & 2 \\

% \midrule
% prec &  \textbf{0.660}          & 0.615         &    0.544       &      \textbf{0.677}     &     0.575      &    0.332       &    \textbf{0.794}      &   0.746        &  0.712 \\
% recall &  0.518          & 0.599         &     \textbf{0.699}      &     0.502      &   0.653        &    \textbf{0.767}       &     0.787                &   0.839        & \textbf{0.871} \\ 
% f1 & 0.581 & 0.607 & \textbf{0.612} & 0.577 & \textbf{0.612} & 0.463 & \textbf{0.791} & 0.79 & 0.783 \\
% tp &  1290          & 1490        &  1740         &    1414       &    1837      &   2160        &  2144         &   2284        &  2372    \\
% fp &  664          & 932         &  1460         &    674       &    1357       &   4356        &  556         &   777        &   962    \\ 
% \bottomrule

% \end{tabularx}

%% file: rq4.tex
%auto-ignore

\smallskip
\RQ{3}{\rqd}
\label{sec:rq4}

We explore this RQ using \toolBB, as discussed in~\Cref{subsec:black_box}. We assume only having access to unlabeled testing data and the model under test as a black-box. 
To evaluate \toolBB, we spatially transform each test input $m$  times by randomly applying $d\omega\in[-30,+30]$ degrees rotation, $dx\in[-3,+3]$ pixels horizontal translation, and $dy\in[-3,+3]$ pixels vertical translation. We then calculate the output diversity score ($\lambda$) based on~\Cref{eq:diversity} and rank the test images based on $\lambda$. 
Finally, we mark top $k$ images as potential most non-robust points. The parameter $k$ is chosen according to users' need.

\begin{wrapfigure}{l}{0.45\textwidth}
\vspace{-3mm}
    \centering
    \includegraphics[width=0.35\textwidth]{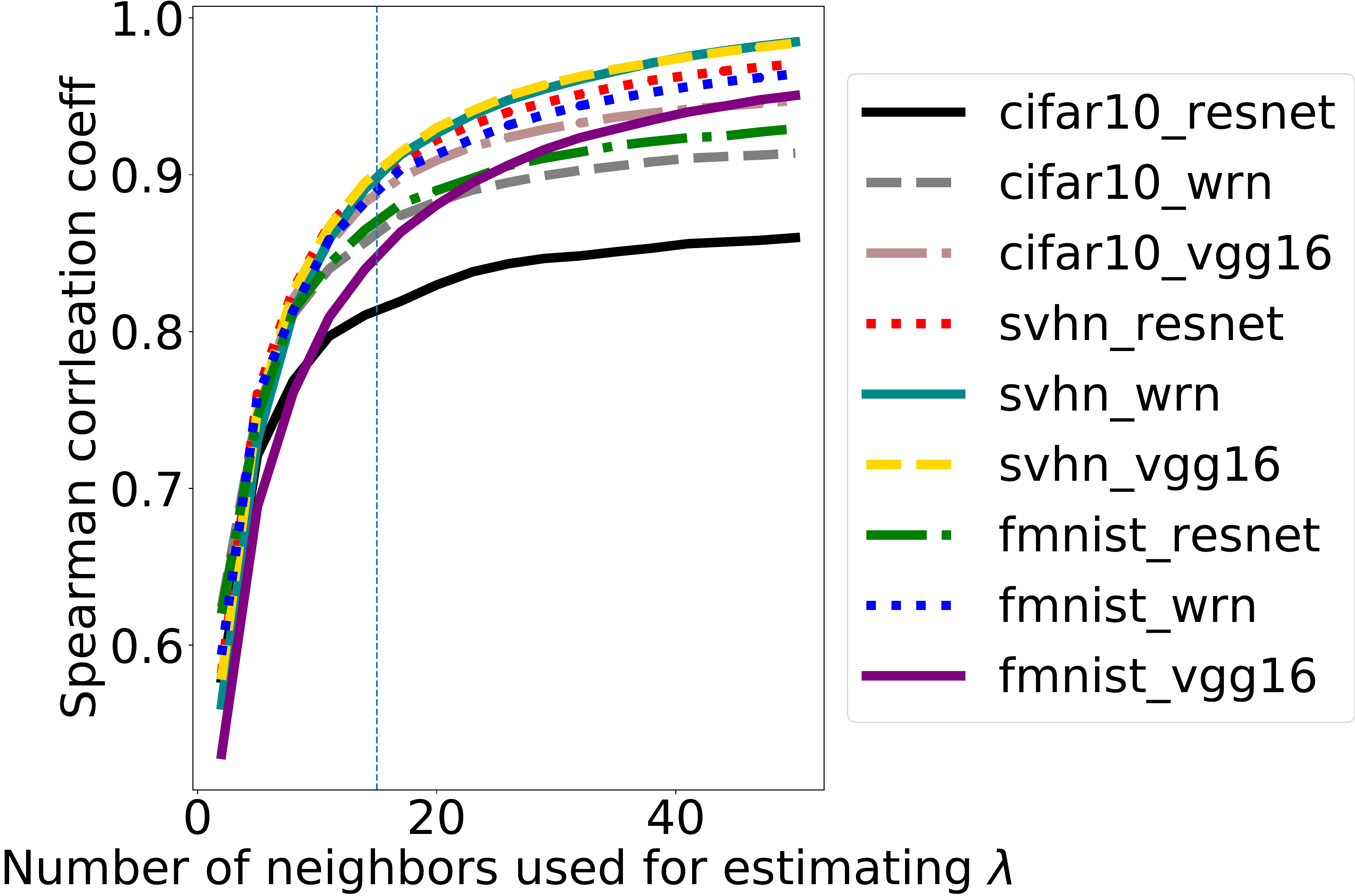}
    \caption{\small{\textbf{The spearman correlation coeff.~between diversity score ($\lambda$) and neighbor accuracy, with varying \#neighbors ($m$).}}}
    \label{fig:spearman_si_neighbor_acc_wrt_num}
% \vspace{-1mm}
\end{wrapfigure}

\begin{wrapfigure}{r}{0.5\columnwidth}
 \vspace{-54mm}
\centering
    \includegraphics[width=0.4\textwidth]{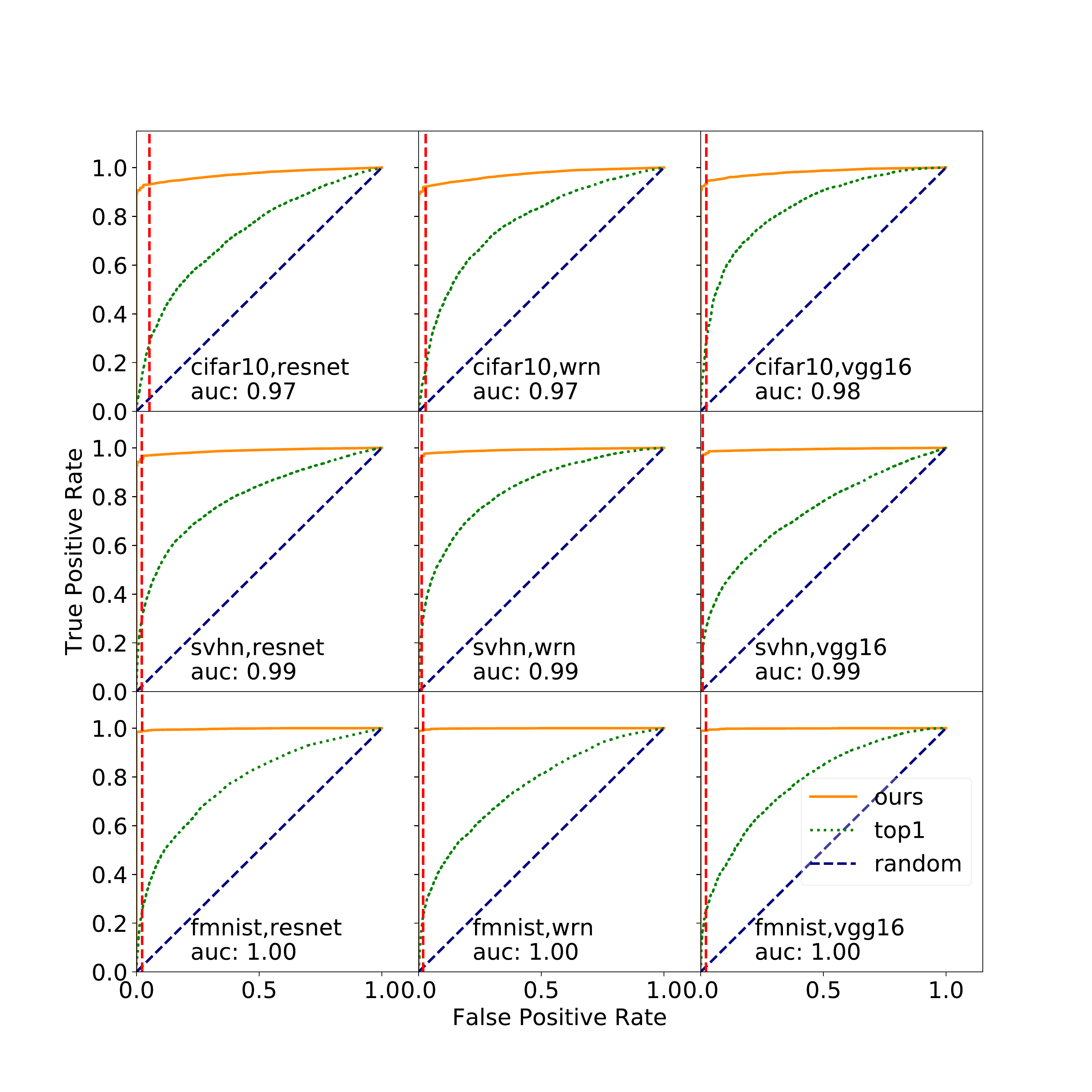}
    \caption{\small{\textbf{AUC-ROC curve with neighbor accuracy cutoff at 0.75. The red vertical line indicates when the diversity score threshold is chosen from training data.}}}
    % \caption{\small{\textbf{AUC-ROC curve with neighbor accuracy cutoff at 0.75. The red vertical line indicates the  false- and true- positive rate when the diversity score threshold is chosen from training data.}}}
    \label{fig:div_roc_auc}
    \vspace{-9mm}
\end{wrapfigure}

With each test data, \toolBB queries the model with $m$ neighbors to compute $\lambda$. Since querying the classifier comes with an overhead, our goal is to achieve an optimal accuracy with minimal queries (\ie $m$). To determine an optimal $m$ value, we explore the spearman correlation between diversity score and neighbor accuracy, with varying $m$, when running \resnet~on all the three datasets (see Figure~\ref{fig:spearman_si_neighbor_acc_wrt_num}). The correlation increases as $m$ increases, as with more query $\lambda$ becomes more accurate, and so the neighbor accuracy. We notice that at $m=15$, the correlation coefficients across all the experimental settings reach above 0.8, and the rate of increase begins to slow down significantly. The results for the other two architectures are highly similar. Thus, we set $m=15$ as default for \toolBB.

% \begin{figure}[!htb]
%   \begin{minipage}{0.48\textwidth}
%     \centering
%     \includegraphics[width=\linewidth]{{"figure/spearmanr_si_neighb_acc"}.pdf}
%     \caption{\small{\textbf{The spearman correlation coeff.~between diversity score ($\lambda$) and neighbor accuracy, with varying \#neighbors ($m$).}}}
%     \label{fig:spearman_si_neighbor_acc_wrt_num}
%   \end{minipage}\hfill
%   \begin{minipage}{0.48\textwidth}
% \centering
%     \includegraphics[width=0.9\textwidth]{figure/div_roc_auc_75.pdf}
%     \caption{\small{\textbf{AUC-ROC curve with neighbor accuracy cutoff at 0.75. The red vertical line indicates the  false- and true- positive rate when the diversity score threshold is chosen from training data.}}}
%     \label{fig:div_roc_auc}
%   \end{minipage}
% \vspace{-10mm}
% \end{figure}

Next, we evaluate \toolBB's performance. We plot AUC-ROC by changing $top-k$ at $m=15$ and compare our method with the random baseline and the top1 baseline as before. 
As shown in Figure~\ref{fig:div_roc_auc}, our method performs much better than the random baseline. In particular, our proposed method achieves AUC higher than 0.87 for all settings when neighbor accuracy cutoff is 0.5 and 0.97 when neighbor accuracy cutoff is 0.75. 
%We also note that if instead of using 15 neighbors but using 50 neighbors to estimate $\lambda$ during the detection time, the performance under each setting will increase by about 0.01\textbackslash 0.06 auc.

Instead of above ranking based scheme, \toolBB can also be used as a classifier if a diversity threshold is given (see~\Cref{subsec:black_box}). 
Here, we estimate the threshold using pre-annotated training data. 
% \toolBB can also be used to classify between strong vs.~ weak point when it is given a diversity score threshold by a user. Such diversity score threshold can be chosen empirically or using an annotated data set. 
% To evaluate \toolBB, we assume the user has access to the training set. 
%In that case, we find the percentile of the chosen neighbor accuracy cutoff and select the corresponding value of diversity score at the same percentile as the diversity score threshold. 
%Similar to the evaluation of \toolWB,

\begin{wraptable}{l}{0.45\columnwidth}
\vspace{-6mm}
\centering
  \footnotesize
  \setlength{\tabcolsep}{3pt}
  \caption{\small{\textbf{Performance of \toolBB and the baseline methods for predicting weak points.}}}
  \label{tab:black_method_classification_performance}
      \resizebox{0.45\columnwidth}{!}{%
    \begin{tabular}{l|l|l|rrr|rrr}
    \toprule
     dataset  & model & method & \multicolumn{3}{c|}{75\%} & \multicolumn{3}{c}{50\%} \\
    \midrule
    
      &   &  & f1 & tp & fp & f1 & tp & fp \\
        \midrule
    \cifar & \resnet & ours & 0.939 & 4714 & 257 & 0.622 & 1454 & 801 \\
         & & top1 & 0.376 & 1218 & 206 & 0.182 & 255 & 120 \\
         & & random & 0.501 & 2516 & 2455 & 0.234 & 549 & 1706 \\
         \cmidrule{2-9}
         & \wrn & ours & 0.938 & 3657 & 171 & 0.585 & 986 & 604 \\
         & & top1 & 0.35 & 889 & 222 & 0.183 & 189 & 90 \\
         & & random & 0.383 & 1494 & 2334 & 0.182 & 307 & 1283 \\
         \cmidrule{2-9}
         & \vgg & ours & 0.945 & 3397 & 148 & 0.682 & 1087 & 390 \\
         & & top1 & 0.439 & 1070 & 153 & 0.266 & 278 & 106 \\
         & & random & 0.36 & 1296 & 2249 & 0.153 & 244 & 1233 \\
        \midrule
    \svhn & \resnet & ours & 0.956 & 8371 & 365 & 0.67 & 1845 & 858 \\
         & & top1 & 0.315 & 1665 & 142 & 0.267 & 452 & 122 \\
         & & random & 0.336 & 2944 & 5792 & 0.102 & 280 & 2423 \\
         \cmidrule{2-9}
         & \wrn & ours & 0.963 & 6827 & 227 & 0.718 & 1602 & 514 \\
         & & top1 & 0.292 & 1238 & 130 & 0.203 & 275 & 85 \\
         & & random & 0.275 & 1950 & 5104 & 0.085 & 191 & 1925 \\
         \cmidrule{2-9}
         & \vgg & ours & 0.976 & 8608 & 144 & 0.779 & 2138 & 454 \\
         & & top1 & 0.172 & 840 & 67 & 0.139 & 221 & 52 \\
         & & random & 0.339 & 2997 & 5755 & 0.102 & 279 & 2313 \\
        \midrule
    \fmnist & \resnet & ours & 0.987 & 6422 & 81 & 0.802 & 2316 & 546 \\
         & & top1 & 0.124 & 428 & 11 & 0.039 & 57 & 7 \\
         & & random & 0.655 & 4265 & 2238 & 0.289 & 835 & 2027 \\
         \cmidrule{2-9}
         & \wrn & ours & 0.989 & 6246 & 70 & 0.857 & 2297 & 360 \\
         & & top1 & 0.144 & 490 & 14 & 0.045 & 63 & 8 \\
         & & random & 0.631 & 3987 & 2329  & 0.274 & 736 & 1921 \\
         \cmidrule{2-9}
         & \vgg & ours & 0.991 & 7078 & 60 & 0.847 & 2393 & 418 \\
         & & top1 & 0.104 & 392 & 5 & 0.028 & 39 & 5 \\
         & & random & 0.711 & 5084 & 2054 & 0.277 & 784 & 2027 \\
  \bottomrule
    \end{tabular}
    }
\vspace{-8mm}
\end{wraptable}

We evaluate precision and recall of \toolBB in the nine DNN-dataset combinations under neighbor accuracy cutoffs 0.5 and 0.75. Table \ref{tab:black_method_classification_performance} shows the result. At $0.75$ setting, \toolBB has f1 up to $99.1\%$, with an average of $96.5\%$.  At $0.50$ setting, \toolBB detects weak points with average f1 of $72.9\%$, while it can go up to $85.7\%$. It consistently produces better estimation than the top1 baseline and the random baseline. This shows that our black-box method can effectively identify weak points.

% We also give precision, recall, the number of true positives and false positives at a particular diversity score cutoff when using training data as discussed in Section~\ref{subsec:black_box}. The corresponding cutoffs are labeled as red, dotted vertical lines in Figure \ref{fig:div_roc_auc}. The comprehensive results are shown in Table \ref{tab:black_method_classification_performance}. Our method has precision as high as $97.8\%$ and recall as high as $97.2\%$ under the $75\%$ setting, precision as high as $85.6\%$ and recall as high as $83.6\%$ under the $50\%$ setting. It consistently produces much better estimation than the random method baseline under all the settings. This shows that our black-box method can also effectively identify weak points.

% \begin{figure*}[htbp]
% \centering

% \subfloat[neighbor accuracy cutoff is 0.5]{\includegraphics[width=0.49\textwidth]{{"figure/div_roc_auc_15_5"}.pdf}}
% \subfloat[neighbor accuracy cutoff is 0.75]{\includegraphics[width=0.49\textwidth]{{"figure/div_roc_auc_15_75"}.pdf}}

% \caption{\textbf{\small{AUC-ROC curve. The red vertical line indicates the  false- and true- positive rate when the diversity score threshold is chosen from training data.}}}
% \label{fig:div_roc_auc}
% \end{figure*}

%It should be noted that compared with DeepRobust white-box, DeepRobust black-box needs extra data transformation and DNN queries in order to calculate SI and thus lead to slightly higher overhead in the real time. 

Note that, generating the spatial transformations and querying the model with it under black box setting is fast. Previous black box methods for adversarial perturbation work in such fashion \cite{pmlr-v97-guo19a, pmlr-v97-moon19a}. % so it is still possible to use this method in real time. 
For example, using \cifar~, when we use a batch with size 100, 
the average transformation+query time for one image is $0.031\pm0.015$ ms. For the other two datasets, the overhead is similar. 
Thus, to for $m=15$ queries, it takes only $0.465\pm0.225$ ms, which is a negligible overhead for most real-world DNN based vision applications. This implies that our black-box method can also be used in real time for many applications.

\smallskip
\RS{3}{Given only black-box access to the DNN classifier, \toolBB can identify weak points with f1 that are much better than those of using top1 method or random method.}

%% file: rq5.tex
\smallskip
\RQ{4}{\rqe}
\label{sec:rq5}

% \begin{figure}[h!tbp]
% % \vspace{-3mm}
% \centering
% \includegraphics[width=0.45\columnwidth]{{"figure/tsne/self_driving_epoch"}.pdf}
% \caption{\textbf{\small{The t-SNE plot of correctly classified data points from the~\sdc~ dataset when using the epoch model. Each data point is colored based on its neighbor accuracy. 
% }}}
% \label{fig:tsne_3d}
% % \vspace{-3mm}
% \end{figure}

% \begin{table}[ht]
% \begin{minipage}[b]{0.56\linewidth}
% \centering
% \begin{tabular}{ | l | r | r | r |}
%     \hline
%     Student & Hours/week & Grade \\ \hline \hline
%     Ada Lovelace & 2 & A \\ \hline
%     Linus Thorvalds & 8 & A \\ \hline
%     Bruce Willis & 12 & F \\ \hline
%     Richard Stallman & 10 & B \\ \hline
%     Grace Hopper & 12 & A \\ \hline
%     Alan Turing & 8 & C \\ \hline
%     Bill Gates & 6 & D \\ \hline
%     Steve Jobs & 4 & E \\ \hline
%   \end{tabular}
%     \caption{Student Database}
%     \label{table:student}
% \end{minipage}\hfill
% \begin{minipage}[b]{0.4\linewidth}
% \centering
% \includegraphics[width=40mm]{example-image}
% \captionof{figure}{2-D scatterplot of the Student Database}
% \label{fig:image}
% \end{minipage}
% \end{table}

%\edited{The local robustness issues also exist in more critical applications like self-driving-car. As shown in \Cref{{fig:tsne_3d}},  %shows a 3D t-SNE plot of a subset of the correctly classified data points from the self-driving car dataset. It shows that 
%even for those correctly classified data points, 
%there is a non-trivial portion of the data points 
%(in the heatmap, more red signified weaker) suffer from 
%low neighbor accuracy.} 

\begin{wrapfigure}{r}{0.45\columnwidth}
    \vspace{-12mm}
 \centering
    \includegraphics[width=0.35\columnwidth]{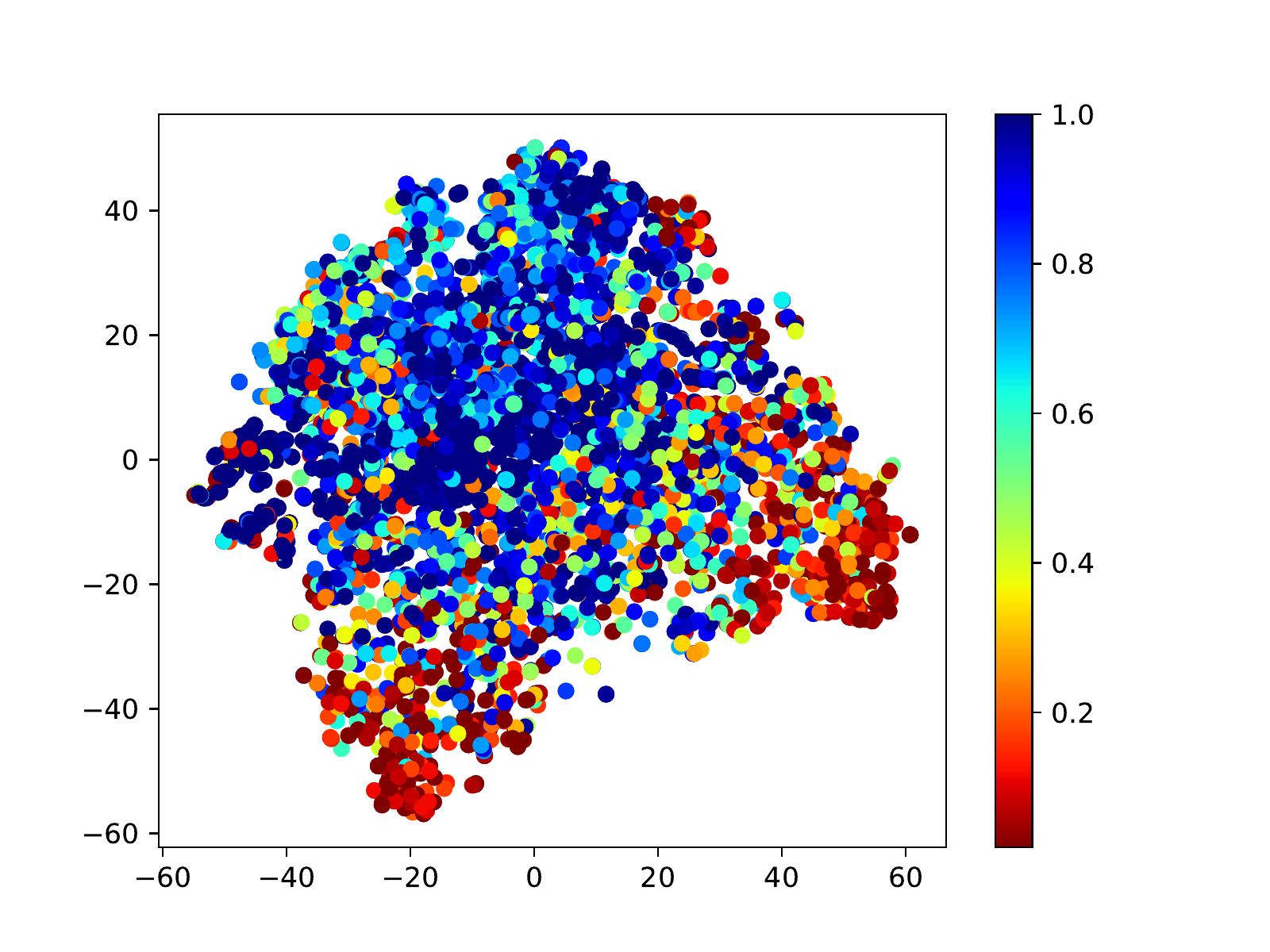}
    
    \captionof{figure}{\textbf{\small{The t-SNE plot of correctly classified data points from~\sdc~ dataset by the epoch model. data points are colored based on neighbor accuracy. 
    }}}
    \label{fig:tsne_3d}
    \vspace{-10mm}
\end{wrapfigure}

The local robustness issues also exist in more critical applications like self-driving-car. Here we explore more complex transformations, \ie adding rain and fog to the driving scenes. As shown in \Cref{fig:tsne_3d}, among those correctly classified data points, there is a non-trivial portion (45.8\%) of them (in the heatmap, more red signified weaker) suffer from 
low (<0.75) neighbor accuracy.

Note that, here, we test regression models, which take images 
of driving scenes as inputs and output the corresponding 
steering angles.

Let a set of outputs predicted by a DNN be denoted by $\{\hat{\theta}_{o1}, \hat{\theta}_{o2}, ... ,\hat{\theta}_{on}\}$, and ground truth labels for the original (unmodified) image points be $\{\theta_1, \theta_2, ..., \theta_n\}$.
If the difference between predicted steering angle $\hat{\theta}_{oi}$ of a transformed image and the ground truth label of the original image $\theta_i$ is above a threshold, we consider it as incorrect. 

The threshold $\lambda MSE_{orig}$ is defined following DeepTest's~\cite{tian2017deeptest},  where $MSE_{orig}=\frac{1}{n} \sum_{i=1}^{n} (\theta_i-\hat{\theta}_{oi})^2$ . 
$MSE$ is the Mean Square Error between the outputs and 
the manual labels, and $\lambda$ is a positive coefficient that 
is chosen to reflect a user's tolerance on the deviation.  
Note that there is no softmax layer (and thus no confidence score) in these regression models so the top1 baseline method cannot be used here. 

Table~\ref{tab:self_driving} shows the result when $\lambda=3$. At $0.75$ setting, \toolWB has f1 score up to $78.9\%$, with an average of $58.2\%$.  At $0.50$ setting, \toolWB detects weak points with an average f1 of $47.9\%$, while it can go as high as $68.2\%$. It consistently produces much better estimation than the random baseline under all the settings. It should be noted that our observation is valid for all the $\lambda$ used in \cite{tian2017deeptest} from $\lambda$ equal to 1 to 5. This shows that our proposed method \toolWB can be applied to regression problems with more complex natural transformations.

\begin{wraptable}{l}{0.45\columnwidth}
 \vspace{-8mm}
  \centering
  \footnotesize
  \setlength{\tabcolsep}{3pt}

  \resizebox{0.45\columnwidth}{!}{%
    \begin{tabular}{l|r|rrr|rrr}
    \toprule
    
         model & method & \multicolumn{3}{c|}{0.75 neighbor acc.} & \multicolumn{3}{c}{0.50 neighbor acc.} \\
    \midrule
    
        &  & f1 & tp & fp & f1 & tp & fp \\
        \midrule
        chauffeur & ours & 0.417 & 555 & 547 & 0.346 & 339 & 384 \\
         & random & 0.146 & 194 & 908 & 0.096 & 94 & 629 \\
           \cmidrule{2-8}
         epoch & ours & 0.789 & 4354 & 1112 & 0.682 & 2641 & 1127\\
         & random & 0.586 & 3234 & 2232 & 0.411 & 1592 & 2176 \\
           \cmidrule{2-8}
          dave2 & ours & 0.541 & 979 & 471 & 0.409 & 475 & 246 \\
         & random & 0.193 & 350 & 1100 & 0.121 & 141 & 580 \\
  \bottomrule
    \end{tabular}
   }
    \caption{\small{\textbf{Performance of \toolWB %and the random baseline method 
  for predicting weak points of \sdc~dataset}}}
  \label{tab:self_driving}

    \vspace{-8mm}
\end{wraptable}

It should also be noted that it is unrealistic to use \toolBB for this task for two reasons: It is impractical to try different variations of an image in real-time for a self-driving car, which is a time-sensitive application. Further, \toolBB requires the calculation of neighbor diversity score. For a regression problem, the predicted values are continuous, so there is a very low probability for any two predictions being equal. Thus, the neighbor diversity score for every data point will be the same and cannot be used for identifying the weak points.
% \smallskip

\RS{4}{\toolWB can detect weak points of a self-driving car dataset with f1 score up to 78.9\%, with an average of 58.2\%, at neighbor accuracy cutoff 0.75.}

%\RS{5}{\toolWB can detect weak points in the regression task of predicting the steering angle of a self-driving car under natural variantions of rain or fog. It can identify non-robust points with f1 up to 78.9\%, with an average of 58.2\%, at neighbor accuracy cutoff 0.75.}

%% file: related_work.tex
%auto-ignore
%!TEX root = main.tex
\section{Related Work}
\label{sec:rel}

\noindent\textbf{Adversarial examples.} Many works focus on generating adversarial examples to fool the DNNs and evaluate their robustness using pixel-based perturbation \cite{goodfellow2014explaining,carlini2017towards, feinman2017detecting,metzen2017detecting, 10.1109/ICSE.2019.00126, kim2019guiding, wang2018formal, WangFormal2018, wang2018mixtrain, gu2014towards,shaham2015understanding,papernot2016distillation, metricforadv, ilyas2019adversarial}. Some other papers~\cite{spatial2018,rotation2019,engstrom2019exploring}, like us, proposed more realistic transformations to generate adversarial examples. In particular, Engstrom et al. \cite{engstrom2019exploring} proposed that a simple rotation and translation can fool a DNN based classifier, and spatial adversarial robustness is orthogonal to $l_p$-bounded adversarial robustness. However, all these works estimate the overall robustness of a DNN based on its aggregated behavior across many data points. In contrast, we analyze the robustness of individual data points under natural variations and propose methods to detect weak/strong points automatically.

\noindent\textbf{DNN testing.} 
Many researchers~\cite{pei2017deepxplore, ma2018deepgauge,sun2018concolic,kim2019guiding,tian2019deepinspect, DeepBillboard2020, DeepImportance2020,Misbehaviour2020,Structure2020, deepfault} proposed techniques to test DNN. For example, Pei et al.~\cite{pei2017deepxplore} proposed an image transformation based differential testing framework, which can detect erroneous behavior by  comparing the outputs of an input image across multiple DNNs. Ferit et al. \cite{deepfault} used fault localization methods to identify suspicious neurons and leveraged those to generate adversarial test cases.

In contrast, others~\cite{tian2017deeptest, zhang2018deeproad, DeepBillboard2020, Structure2020, taejoo19, Callisto, NatPerV} used metamorphic testing where the assumption is the outputs of an original and its transformed image will be the same under natural transformations. Among them, some use a uncertainty measure to quantify some types of non-robustness of an input for prioritizing samples for testing / retraining \cite{taejoo19} or generating test cases\cite{Callisto}. We follow a similar metamorphic property while estimating neighbor accuracy and our proposed \toolBB also leverages an uncertainty measure. The key differences are: First, we focus on estimating model's performance on general natural variants of an input rather than the input itself or only spatial variants. Second, we focus on the task of weak points detection rather than prioritizing / generating test cases. We also give detailed analyses of the properties of natural variants and propose a feature vector based white-box detection method \toolWB. Further, we show that our method works across domains (both image classification and self-driving car controllers) and tasks (both classification and regression). Other uncertainty work complement ours in the sense that we can easily leverage weak points identified by \toolWB and \toolBB to prioritize test cases or generate more adversarial cases of natural variants.

Another line of work \cite{Qiu2020Quantifying, pmlr-v48-gal16, pmlr-v70-guo17a, pmlr-v80-teye18a, attribution_conf, trust_score, UncertaintyDL} estimates the confidence of a DNN's output. For example, \cite{pmlr-v48-gal16} leverages thrown away information from existing models to measure confidence; \cite{pmlr-v70-guo17a} shows other NN properties like depth, width, weight decay, and batch normalization are important factors influencing prediction confidence. Although such methods can provide a confidence measure per input or its adversarial variants, they do not check its natural robustness property, i.e., with natural variations how will they behave. 

\noindent\textbf{DNN verification.} There also exist work on verifying properties for a DNN model \cite{pei2017towards,wang2018formal,huang2017safety,seshia2018formal,ehlers2017formal, piecewiseLNNV, RobustVEnsemble}. Most of them focus on verifying properties on $l_p$ norm bounded input space. Recently, Balunovic et al.\cite{geometric_certify} provides the first verification technique for verifying a data point's robustness against spatial transformation. However, their technique suffers from scalability issues.

\noindent\textbf{Robust training.}
Regular neural network training involves the optimization of the loss value for each data point. 
Robust training of neural network works on minimizing the largest loss within a specific bounded region usually using adversarial examples~\cite{wang2018mixtrain,katz2017reluplex,mirman2018differentiable,wong2018scaling,wang2018formal,rotation2019,madry2018,tramer2017ensemble,ma2018characterizing}.
While both robust training methods and our work generate variants of data points, instead of training a model with these variants to improve robustness, we leverage them to estimate the robustness of the unseen data points. The relation between robust retraining and our work is thus similar to bug fixing vs.~bug detection in traditional software engineering literature. 

%% file: threats.tex
\section{Threats to Validity}
\label{sec:threats}
% First, there are several different threat models and corresponding definitions of robustness e.g. norm-bounded robustness. It will be interesting to check if our observations and detection methods can be directly used or adapted for the other threat models and non-spatial transformations. 

% First, it will be interesting to explore what makes a small portion of weak points lie closer to the class center in the representation space. It also worth exploring why some points have large neighbor accuracy change across different models.

% Second, there is a large space to explore to improve the white-box detector's performance on identifying weak points. For example, one may try to use more complex structures e.g. CNN, \resnet, DenseNet or some ensemble models. The hyper-parameters (optimization method, learning rate schedule, training epoch) can also be fine-tuned further. 

% Last but not least, although our approach is conceptually general, it may be interesting to see whether our approach can be directly applied or adapted to the other domains such as speech recognition, machine translation, and so on.

%We only test on 12 dataset-model combinations. However, the datasets and models we choose are the very popular ones. 
We adopt rotation and translation as transformations for image classification tasks and rain and fog effects for the self-driving car task. There are many more natural variations such as brightness, snow effect etc. However, rotation and translation are representative of spatial transformation and used by many paper in evaluating robustness of DNN models\cite{engstrom2019exploring,pei2017deepxplore}. Rain and fog effects are also widely leveraged in many influential studies on testing self-driving cars \cite{pei2017deepxplore,tian2017deeptest,zhang2018deeproad}.

Besides, for some of the experiments we did not show all the combinations under both neighbor accuracy cutoffs (i.e. 0.5 and 0.75). However, we note that the observations are consistent and we did not include them purely because of space limitation. Another limitation is that for both \toolWB and \toolBB, we need to decide the number of neighbors to use for training a classifier and estimating $\lambda$, respectively. We mitigate it by selecting the neighbor numbers that give stable performance in terms of precision and recall. %Besides, our study on spatial robustness only consider the rotation and translation transformations. However, these two are the most popular ones in the literature studying spatial robustness.

%% file: conclusion.tex
%auto-ignore
\section{Conclusion and Future Work}
\label{sec:conclusion}
% In this work, we involve the data characteristic into the robustness testing of a DNN model. We adopt the concept of neighbor accuracy as a measure for local robustness of a data point on a given model. We explore the properties of neighbor accuracy and find that it is a relatively stable measure across models for most data points and thus, inherent to the dataset rather than the model used. We also observe that weak points are often located towards the corresponding class boundaries and their transformed versions are likely to be predicted to be more diverse classes. By leveraging these two observations, we propose a white-box method and a black-box method to identify weak/strong points to warn a user about potential weakness in the given trained model in real-time. We design, implement and evaluate our proposed framework, DeepRobust, on three image recognition datasets and one self-driving car dataset with three models for each. The results show that DeepRobust is able to identify weak points %(and thus vulnerable regions) 
% and strong points %(and thus robust regions) 
% with high precision and recall.  
% Besides, it allows a user to trade off precision for recall to accommodate different needs.

In this work, we involve the data characteristic into the robustness testing of DNN models. We adopt the concept of neighbor accuracy as a measure for local robustness of a data point on a given model. We explore the properties of neighbor accuracy and find that weak points are often located towards corresponding class boundaries and their transformed versions tend to be predicted to be more diverse classes. Leveraging these observations, we propose a white-box method and a black-box method to identify weak/strong points to warn a user about potential weakness in the given trained model in real-time. We design, implement and evaluate our proposed framework, \toolWB and \toolBB, on three image recognition datasets and one self-driving car dataset (for \toolWB only) with three models for each. The results show that they can effectively identify weak/strong points with high precision and recall. 

% Besides, it allows a user to trade off precision for recall to accommodate different needs.

For future work, other consistency analysis methods \cite{UncertaintyDL} e.g. variation ratio, entropy can be tried. \edited{We can potentially attain statistical guarantee for our black-box method by modeling the neighbor accuracy distribution and assume certain level of correlation between neighbor accuracy and complexity score. Besides, other definitions of robustness like consistency can be explored.} We can also leverage ideas from \cite{taejoo19, Callisto} to easily prioritize test cases or generate more hard test cases based on identified weak points. Further, we can potentially modify existing fixing methods such as \cite{Sensei:ICSE2020} targeting the weak points to fix them.

% First, it will be interesting to explore what makes a small portion of weak points lie closer to the class center in the representation space. It also worth exploring why some points have large neighbor accuracy change across different models.

% Second, there is a large space to explore to improve the white-box detector's performance on identifying weak points. For example, one may try to use more complex structures e.g. CNN, \resnet, DenseNet or some ensemble models. The hyper-parameters (optimization method, learning rate schedule, training epoch) can also be fine-tuned further. 

% Last but not least, although our approach is conceptually general, it may be interesting to see whether our approach can be directly applied or adapted to the other domains such as speech recognition, machine translation, and so on.

%% file: acknowledgement.tex
\section{Acknowledgement}
\edited{ We thank Mukul Prasad and Ripon Saha from Fujisu US for valuable discussions. This work is supported in part by NSF CCF-1845893 and CCF-1822965.}

%% file: appendix.tex
%auto-ignore

{\noindent\large\bf Appendix}
\section{More Empirical Results on Characteristics of Spatial Robustness}
\label{appendix}
\noindent
We first explore the distribution of spatial robustness %, measured using  {\em neighbor accuracy}, 
of the studied data sets w.r.t. three well-trained models.
%We measure spatial robustness in terms of {\em neighbor accuracy}. 
In particular, we explore the variations of spatial robustness: a) across different data points of a given model, b) across different models of a given data point.

% \renewcommand{\theenumi}{\alph{enumi}}
% %\begin{itemize}
% \begin{enumerate}[label=\alph*),leftmargin=\parindent] 
%     \item  across different data points of a given model. 
%     \item across different models of a given data point.
%     %\item Given a well trained model, how spatial robustness changes across different data points?
%   % \item RQ1b. Across different models, does spatial robustness of a given point vary?
%   % \item Given a data point, how spatial robustness changes across different models?

% \end{enumerate}

\begin{figure}[!htpb]
\centering
       \includegraphics[width=0.6\linewidth]{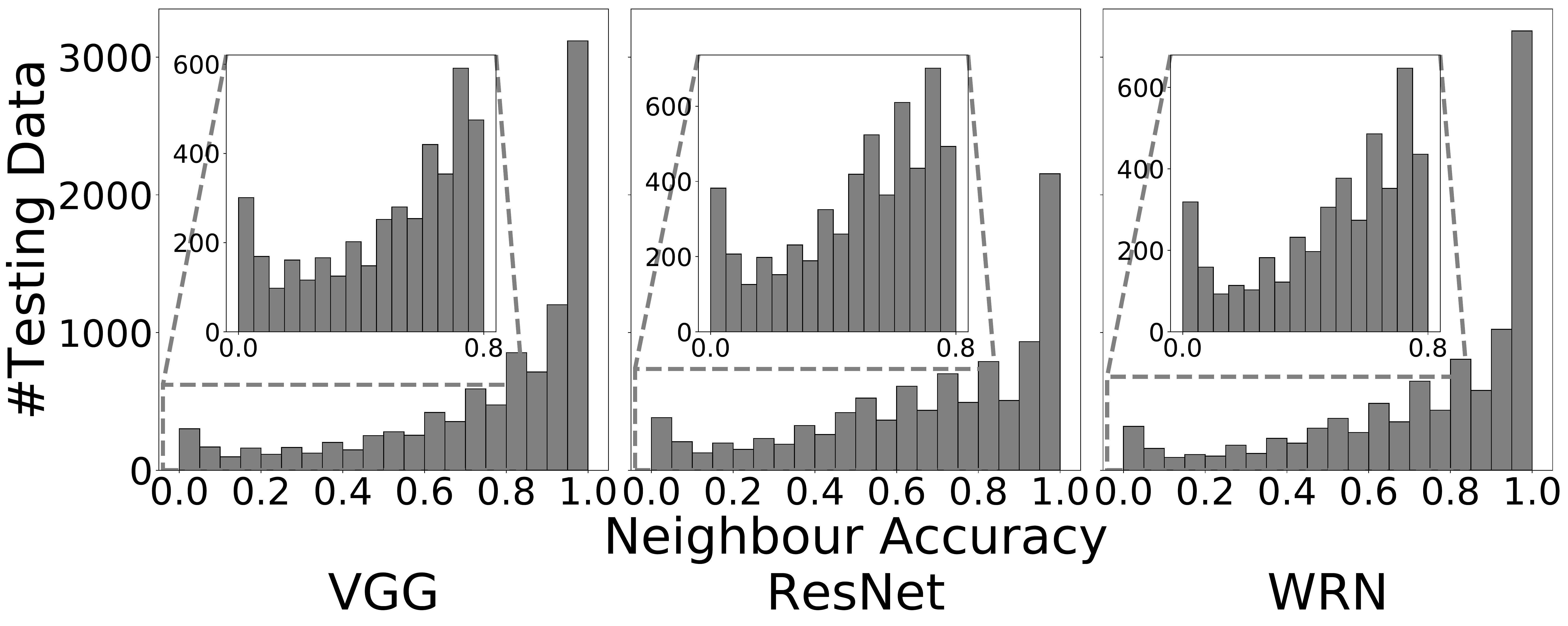}
  %\caption{\small Train data neighbor accuracy}
\caption{\textbf{\small Histogram of neighbor accuracy for \cifar}}
\label{fig:histogram1}
% \vspace{-0.3cm}
\end{figure}

% \begin{figure}[!htpb]
% \centering
      
%   \includegraphics[width=\linewidth]{experiments/rq1/cifar10_2_withoutlastbin.pdf}
%   %\caption{\small Train data neighbor accuracy}
% \caption{\textbf{\small Histogram of neighbor accuracy for \cifar~without last bin}}
% \label{fig:histogram2}
% %\vspace{-0.3cm}
% \end{figure}

%\RQ{1a}{How neighbor accuracy of data points in a dataset is distributed on a well trained model?}
\medskip
\RQA{ appendix 1a}{Given a well trained model, how spatial robustness changes across different data points?} 
%Previous work tends to focus on an aggregated spatial robust accuracy measure, i.e., evaluating the overall accuracy of a model under different transformations \cite{engstrom2019exploring}. 
Here, we explore the spatial robustness of each point as captured by its neighbor accuracy. To have a fine-granular understanding, we draw the distribution of neighbor accuracy of data points in both training set and testing set of \cifar\ when using \vgg, \resnet\, and \wrn\  models, respectively, as shown in Figure \ref{fig:histogram1}. The histograms show that for all these models, the neighbor accuracy vary widely across data points\textemdash most of the data points have very high neighbor accuracy, %(60\% of training data and 58\% testing data have neighbor accuracy $>=$0.75),  
there is a non-trivial number of points having relatively low neighbor accuracy (40\% of training data and 42\% of testing data have neighbor accuracy $<$0.75). 
%For example, we see for \vgg\ model, 17.8\% of the testing data points have less than 50\% neighbor accuracy; 
We call such low accuracy points to be  {\em weak} or {non-robust points}. 
Similarly, for neighbor accuracy $<$0.50, around 16\%  of training data and 20\% of testing data turned out to be weak. 
These points degrade the aggregated spatial robustness of the model. The same finding also holds for the other two datasets.

The distribution of neighbor accuracy on the training data (not shown in Figure \ref{fig:histogram1}) looks similar to that for the testing data. Table \ref{tab:correlation2} 
further shows that correlation of the number of points in each neighbor accuracy bin between training and testing data is very strong (0.999 to 0.944) with statistical significance ($p < 0.05$). %Thus, a strong correlation exists between the training data and testing data in terms of neighbor accuracy. 
%Even among less robust points (neighbor accuracy $<$ 1) this observation is true. 
%if we do not consider perfectly robust points (as spatial robustness is imbalanced), we see high correlation between training and testing data (the last  row of  the table). 
Such strong correlation is perhaps not surprising, as training and testing data are taken from the same data distributions.

{
\begin{table}[htpb]
\setlength{\tabcolsep}{2pt}
\centering
\footnotesize
\vspace{-3mm}
\caption{\small{\textbf{Correlation of spatial robustness (\ie the number of points per neighbor accuracy bin) between training and testing data}}}
\label{tab:correlation2}
\begin{tabular}{l|l|l|l|l|l|l|l|l|l}
\toprule
Dataset  & \multicolumn{3}{c|}{\cifar} & \multicolumn{3}{c|}{\svhn} & \multicolumn{3}{c}{\fmnist} \\
\midrule
Model  & \vgg & \resnet & \wrn & \vgg & \resnet & \wrn & \vgg & \resnet & \wrn \\

\midrule
spearman &  0.944          & 0.979         &    0.970       &     0.977      &     0.997      &    0.998       &    0.995      &   0.983        &  0.995 \\
pearson  &  0.998          & 0.996         &    0.997       &     0.998      &     0.995      &    0.999       &    0.995      &   0.990        &  0.995  \\ 
%pearson* &  0.990          & 0.965         &  0.973         &    0.963       &    0.960       &   0.964        &  0.916         &   0.879        &   0.921    \\ 
\bottomrule
%\multicolumn{10}{l}{* w/o last bin}
%\vspace{-3mm}
\end{tabular}
\end{table}

% \RS{1c}{Training and testing data tend to have similar neighbor accuracy distribution.}
}

% \RS{1a}{
% A non-trivial number of data points across all the datasets are weak \textemdash around 46\% (at cutoff $<$ 0.75) and 18\%  (at cutoff $<$ 0.50) points, on average, have low neighbor accuracy making them non-robust under spatial transformations. %Moreover, while the robustness seems to be an intrinsic property of the data, i.e., a non-robust data point is non-robust across multiple models, 2\%(average all setting with change smaller than 0.5)  of the data points' robustness is also dependent on the models.
% }

\begin{figure}[!htpb]
%\vspace{-3mm}
\centering
          \includegraphics[width=0.6\linewidth]{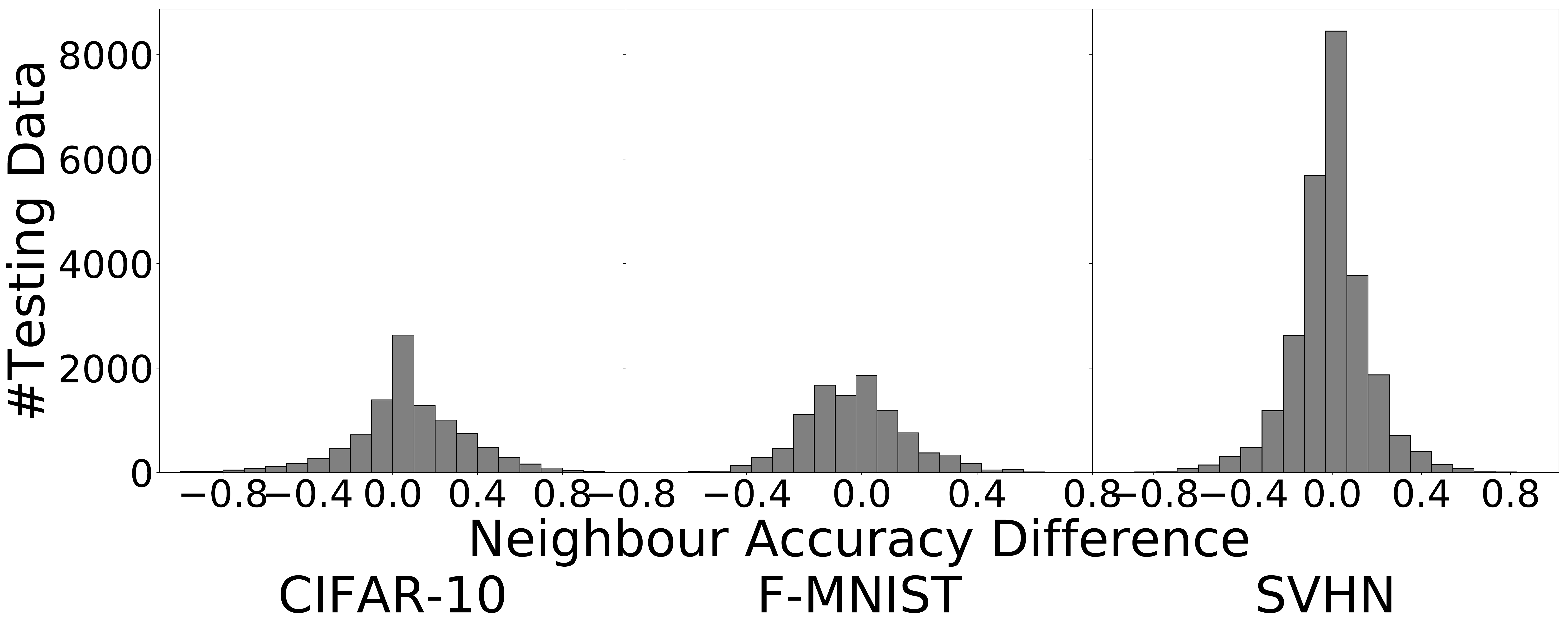}
          %\caption{\small Per datapoint neighbor accuracy change}
\caption{\textbf{\small Histogram of per data point neighbor accuracy changes for three datasets from \vgg~ model to \resnet~ model across all the three datasets on testing data. }}
\label{fig:histogram3}
%\vspace{-0.3cm}
% \vspace{-3mm}
\end{figure}

\RQA{ appendix 1b}{Given a data point, how spatial robustness changes across different models?}
Figure \ref{fig:histogram1} shows that the distribution of neighbor accuracy for a dataset looks similar across different models, indicating robustness may be a property of the dataset itself and may not depend on the underlying model architecture. To dig further, for a given dataset, we divide the neighbor accuracy of all data points into 20 equal-sized bins and compute the Spearman and Pearson correlations of the number of points falling into every bin between any two models.   To reduce the potential bias induced by the large portion of original points having neighbor accuracy higher than 95\%, we also compute the Pearson correlation without them. Table~\ref{tab:correlation1} shows that the correlations are very high between any two models for all the three datasets confirming that the overall robustness distribution of a dataset is very similar across different models.

\begin{table}[htpb]
% \vspace{-3mm}
\setlength{\tabcolsep}{3pt}
\centering
\footnotesize
\caption{\small{\textbf{Correlation between models in number of points per neighbor accuracy bin}}}
\label{tab:correlation1}
%\begin{tabularx}{0.95\columnwidth}{l|l|l|c|c|c}
\begin{tabular}{l|l|l|r|r|r}
\toprule
Dataset  &  Setting & Data & Spearman & Pearson & Pearson \\ 
  &   &  &  &  & (w/o last bin) \\ \midrule
\multirow{6}{*}{\cifar} & \multirow{2}{*}{\vgg\ vs \resnet}& Train &0.980&0.985&0.938 \\ %\cmidrule(l){3-6}
& & Test&0.956&0.980&0.920 \\ \cmidrule(l){2-6} 

& \multirow{2}{*}{\vgg\ vs \wrn}&Train&0.995&0.996&0.986 \\ %\cmidrule(l){3-6}
& &Test&0.977&0.995&0.980 \\ \cmidrule(l){2-6}
& \multirow{2}{*}{\resnet\ vs \wrn}&Train&0.991&0.990&0.980 \\ %\cmidrule(l){3-6}
& &Test&0.989&0.986&0.977 \\ \midrule
\multirow{6}{*}{\svhn} & \multirow{2}{*}{\vgg\ vs \resnet}&Train&0.955&0.978&0.979 \\ %\cmidrule(l){3-6}
& &Test&0.995&0.983&0.995 \\ \cmidrule(l){2-6}
& \multirow{2}{*}{\vgg\ vs \wrn}&Train&0.977&0.979&0.993 \\ %\cmidrule(l){3-6}
& &Test&0.998&0.983&0.994 \\ \cmidrule(l){2-6}
& \multirow{2}{*}{\resnet\ vs \wrn}&Train&0.992&0.986&0.955 \\ %\cmidrule(l){3-6}
& &Test&0.998&0.995&0.981 \\ \midrule
\multirow{6}{*}{\fmnist} & \multirow{2}{*}{\vgg\ vs \resnet}&Train&0.863&0.637&0.969 \\ %\cmidrule(l){3-6}
& &Test&0.890&0.627&0.970 \\ \cmidrule(l){2-6}
& \multirow{2}{*}{\vgg\ vs \wrn}&Train&0.877&0.883&0.986 \\ %\cmidrule(l){3-6}
& &Test&0.893&0.874&0.969 \\ \cmidrule(l){2-6}
& \multirow{2}{*}{\resnet\ vs \wrn}&Train&0.967&0.901&0.965 \\ %\cmidrule(l){3-6}
& &Test&0.979&0.899&0.979 \\ %\cmidrule(l){2-6}
\bottomrule
%\vspace{-3mm}
\end{tabular}
%\end{tabularx}
\end{table}

We further examine the per-point variation of robustness across models by analyzing how neighbor accuracy of each data point varies across models.  Figure \ref{fig:histogram3} shows that the change is minimal:  for \cifar, \fmnist~and \svhn, $60\%$, $76\%$, and $81\%$ of data points have neighbor accuracy change less than $0.2$ across any two models. This implies that a large portion of data points' neighbor accuracy is independent of the model selected. 
Note that data points that are non-robust across all the models will not be identified by differential testing techniques as proposed by DeepXplore~\cite{pei2017deepxplore}.

% \RS{1b}{Most data points (75\%) have similar neighbor accuracy, aka robustness,  across different models indicating natural robustness is a property of data than model. }
% \smallskip
\RS{appendix 1}{A non-trivial number of data points (46\% at neighbor accuracy $<0.75$) of the whole data set are weak. 
Robustness seems to be an intrinsic property of the data points and does not vary across models.
}

%% file: main_fase.bbl
\begin{thebibliography}{10}
\providecommand{\url}[1]{\texttt{#1}}
\providecommand{\urlprefix}{URL }
\providecommand{\doi}[1]{https://doi.org/#1}

\bibitem{chauffeur}
Chauffeur model.
  \url{https://github.com/udacity/self-driving-car/tree/master/steering-models/community-models/chauffeur}
  (2016)

\bibitem{epoch}
Epoch model.
  \url{https://github.com/udacity/self-driving-car/tree/master/steering-models/community-models/cg23}
  (2016)

\bibitem{MSR-ICSE-SEIP.2019}
Amershi, S., Begel, A., Bird, C., DeLine, R., Gall, H., Kamar, E., Nagappan,
  N., Nushi, B., Zimmermann, T.: Software engineering for machine learning: A
  case study. In: Proceedings of the 41st International Conference on Software
  Engineering: Software Engineering in Practice. pp. 291--300. ICSE-SEIP '19,
  IEEE Press (2019). \doi{10.1109/ICSE-SEIP.2019.00042},
  \url{https://doi.org/10.1109/ICSE-SEIP.2019.00042}

\bibitem{geometric_certify}
Balunovic, M., Baader, M., Singh, G., Gehr, T., Vechev, M.: Certifying
  geometric robustness of neural networks. In: Advances in Neural Information
  Processing Systems. pp. 15287--15297 (2019)

\bibitem{bojarski2016end}
Bojarski, M., Del~Testa, D., Dworakowski, D., Firner, B., Flepp, B., Goyal, P.,
  Jackel, L.D., Monfort, M., Muller, U., Zhang, J., et~al.: End to end learning
  for self-driving cars. arXiv preprint arXiv:1604.07316  (2016)

\bibitem{nvidia-dave2}
Bojarski, M., Testa, D.D., Dworakowski, D., Firner, B., Flepp, B., Goyal, P.,
  Jackel, L.D., Monfort, M., Muller, U., Zhang, J., Zhang, X., Zhao, J., Zieba,
  K.: End to end learning for self-driving cars. CoRR  \textbf{abs/1604.07316}
  (2016), \url{http://arxiv.org/abs/1604.07316}

\bibitem{piecewiseLNNV}
Bunel, R., Turkaslan, I., Torr, P.H., Kohli, P., Kumar, M.P.: A unified view of
  piecewise linear neural network verification. In: Proceedings of the 32nd
  International Conference on Neural Information Processing Systems. p.
  4795–4804. NIPS'18, Curran Associates Inc., Red Hook, NY, USA (2018)

\bibitem{taejoo19}
Byun, T., Sharma, V., Vijayakumar, A., Rayadurgam, S., Cofer, D.: Input
  prioritization for testing neural networks (01 2019)

\bibitem{carlini2017towards}
Carlini, N., Wagner, D.: Towards evaluating the robustness of neural networks.
  In: Security and Privacy (SP), 2017 IEEE Symposium on. pp. 39--57. IEEE
  (2017)

\bibitem{cohen1988spa}
Cohen, J.: {Statistical Power Analysis for the Behavioral Sciences}. Lawrence
  Erlbaum Associates (1988)

\bibitem{DeepStellar:FSE2019}
Du, X., Xie, X., Li, Y., Ma, L., Liu, Y., Zhao, J.: Deepstellar: Model-based
  quantitative analysis of stateful deep learning systems. In: Proceedings of
  the 2019 27th ACM Joint Meeting on European Software Engineering Conference
  and Symposium on the Foundations of Software Engineering. p. 477–487.
  ESEC/FSE 2019, Association for Computing Machinery, New York, NY, USA (2019).
  \doi{10.1145/3338906.3338954}, \url{https://doi.org/10.1145/3338906.3338954}

\bibitem{ehlers2017formal}
Ehlers, R.: Formal verification of piece-wise linear feed-forward neural
  networks. In: International Symposium on Automated Technology for
  Verification and Analysis. pp. 269--286. Springer (2017)

\bibitem{engstrom2017rotation}
Engstrom, L., Tran, B., Tsipras, D., Schmidt, L., Madry, A.: A rotation and a
  translation suffice: Fooling cnns with simple transformations. arXiv preprint
  arXiv:1712.02779  (2017)

\bibitem{engstrom2019exploring}
Engstrom, L., Tran, B., Tsipras, D., Schmidt, L., Madry, A.: Exploring the
  landscape of spatial robustness. In: International Conference on Machine
  Learning. pp. 1802--1811 (2019)

\bibitem{rotation2019}
Engstrom, L., Tran, B., Tsipras, D., Schmidt, L., Mądry, A.: A rotation and a
  translation suffice: Fooling cnns with simple transformations. In:
  Proceedings of the 36th international conference on machine learning (ICML)
  (2019)

\bibitem{deepfault}
Eniser, H.F., Gerasimou, S., Sen, A.: Deepfault: Fault localization for deep
  neural networks. In: H{\"a}hnle, R., van~der Aalst, W. (eds.) Fundamental
  Approaches to Software Engineering. pp. 171--191. Springer International
  Publishing, Cham (2019)

\bibitem{feinman2017detecting}
Feinman, R., Curtin, R.R., Shintre, S., Gardner, A.B.: Detecting adversarial
  samples from artifacts. arXiv preprint arXiv:1703.00410  (2017)

\bibitem{UncertaintyDL}
Gal, Y.: {Uncertainty in Deep Learning}  (2016)

\bibitem{pmlr-v48-gal16}
Gal, Y., Ghahramani, Z.: Dropout as a bayesian approximation: Representing
  model uncertainty in deep learning. In: Balcan, M.F., Weinberger, K.Q. (eds.)
  Proceedings of The 33rd International Conference on Machine Learning.
  Proceedings of Machine Learning Research, vol.~48, pp. 1050--1059. PMLR, New
  York, New York, USA (20--22 Jun 2016),
  \url{http://proceedings.mlr.press/v48/gal16.html}

\bibitem{Sensei:ICSE2020}
Gao, X., Saha, R., Prasad, M., Roychoudhury, A.: Fuzz testing based data
  augmentation to improve robustness of deep neural networks. In: Proceedings
  of the 42nd International Conference on Software Engineering. ICSE 2020, ACM
  (2020)

\bibitem{DeepImportance2020}
Gerasimou, S., Eniser, H.F., Sen, A., Çakan, A.: Importance-driven deep
  learning system testing. In: International Conference of Software Engineering
  (ICSE) (2020)

\bibitem{goodfellow2014generative}
Goodfellow, I., Pouget-Abadie, J., Mirza, M., Xu, B., Warde-Farley, D., Ozair,
  S., Courville, A., Bengio, Y.: Generative adversarial nets. In: Advances in
  neural information processing systems. pp. 2672--2680 (2014)

\bibitem{goodfellow2014explaining}
Goodfellow, I.J., Shlens, J., Szegedy, C.: Explaining and harnessing
  adversarial examples. In: International Conference on Learning
  Representations (ICLR) (2015)

\bibitem{RobustVEnsemble}
Gross, D., Jansen, N., P{\'e}rez, G.A., Raaijmakers, S.: Robustness
  verification for classifier ensembles. In: Hung, D.V., Sokolsky, O. (eds.)
  Automated Technology for Verification and Analysis. pp. 271--287. Springer
  International Publishing, Cham (2020)

\bibitem{gu2014towards}
Gu, S., Rigazio, L.: Towards deep neural network architectures robust to
  adversarial examples. In: International Conference on Learning
  Representations (ICLR) (2015)

\bibitem{pmlr-v97-guo19a}
Guo, C., Gardner, J., You, Y., Wilson, A.G., Weinberger, K.: Simple black-box
  adversarial attacks. In: Chaudhuri, K., Salakhutdinov, R. (eds.) Proceedings
  of the 36th International Conference on Machine Learning. Proceedings of
  Machine Learning Research, vol.~97, pp. 2484--2493. PMLR, Long Beach,
  California, USA (09--15 Jun 2019),
  \url{http://proceedings.mlr.press/v97/guo19a.html}

\bibitem{pmlr-v70-guo17a}
Guo, C., Pleiss, G., Sun, Y., Weinberger, K.Q.: On calibration of modern neural
  networks. In: Precup, D., Teh, Y.W. (eds.) Proceedings of the 34th
  International Conference on Machine Learning. Proceedings of Machine Learning
  Research, vol.~70, pp. 1321--1330. PMLR, International Convention Centre,
  Sydney, Australia (06--11 Aug 2017),
  \url{http://proceedings.mlr.press/v70/guo17a.html}

\bibitem{he2016deep}
He, K., Zhang, X., Ren, S., Sun, J.: Deep residual learning for image
  recognition. In: Proceedings of the IEEE conference on computer vision and
  pattern recognition. pp. 770--778 (2016)

\bibitem{Structure2020}
He, P., Meister, C., Su, Z.: Structure-invariant testing for machine
  translation. In: International Conference of Software Engineering (ICSE)
  (2020)

\bibitem{huang2017safety}
Huang, X., Kwiatkowska, M., Wang, S., Wu, M.: Safety verification of deep
  neural networks. In: International Conference on Computer Aided Verification.
  pp. 3--29. Springer (2017)

\bibitem{ilyas2019adversarial}
Ilyas, A., Santurkar, S., Tsipras, D., Engstrom, L., Tran, B., Madry, A.:
  Adversarial examples are not bugs, they are features (2019),
  \url{http://arxiv.org/abs/1905.02175}

\bibitem{Islam:FSE2019}
Islam, M.J., Nguyen, G., Pan, R., Rajan, H.: A comprehensive study on deep
  learning bug characteristics. In: Proceedings of the 2019 27th ACM Joint
  Meeting on European Software Engineering Conference and Symposium on the
  Foundations of Software Engineering. pp. 510--520. ESEC/FSE 2019, Association
  for Computing Machinery, New York, NY, USA (2019).
  \doi{10.1145/3338906.3338955}, \url{https://doi.org/10.1145/3338906.3338955}

\bibitem{attribution_conf}
Jha, S., Raj, S., Fernandes, S., Jha, S.K., Jha, S., Jalaian, B., Verma, G.,
  Swami, A.: Attribution-based confidence metric for deep neural networks. In:
  Advances in Neural Information Processing Systems. pp. 11826--11837 (2019)

\bibitem{trust_score}
Jiang, H., Kim, B., Gupta, M.: To trust or not to trust a classifier. In:
  Advances in Neural Information Processing Systems. pp. 5541–--5552 (2018)

\bibitem{katz2017reluplex}
Katz, G., Barrett, C., Dill, D.L., Julian, K., Kochenderfer, M.J.: Reluplex: An
  Efficient SMT Solver for Verifying Deep Neural Networks, pp. 97--117.
  Springer International Publishing, Cham (2017)

\bibitem{kim2019guiding}
Kim, J., Feldt, R., Yoo, S.: Guiding deep learning system testing using
  surprise adequacy. In: Proceedings of the 41st International Conference on
  Software Engineering. pp. 1039--1049. IEEE Press (2019)

\bibitem{cifar10}
Krizhevsky, A.: Learning multiple layers of features from tiny images.
  University of Toronto  (05 2012)

\bibitem{krizhevsky2012imagenet}
Krizhevsky, A., Sutskever, I., Hinton, G.E.: Imagenet classification with deep
  convolutional neural networks. In: Advances in neural information processing
  systems. pp. 1097--1105 (2012)

\bibitem{kurakin2016adversarial}
Kurakin, A., Goodfellow, I., Bengio, S.: Adversarial examples in the physical
  world. arXiv preprint arXiv:1607.02533  (2016)

\bibitem{Li:FSE2019}
Li, Z., Ma, X., Xu, C., Cao, C., Xu, J., L\"{u}, J.: Boosting operational dnn
  testing efficiency through conditioning. In: Proceedings of the 2019 27th ACM
  Joint Meeting on European Software Engineering Conference and Symposium on
  the Foundations of Software Engineering. p. 499–509. ESEC/FSE 2019,
  Association for Computing Machinery, New York, NY, USA (2019).
  \doi{10.1145/3338906.3338930}, \url{https://doi.org/10.1145/3338906.3338930}

\bibitem{ma2018deepgauge}
Ma, L., Juefei-Xu, F., Sun, J., Chen, C., Su, T., Zhang, F., Xue, M., Li, B.,
  Li, L., Liu, Y., et~al.: Deepgauge: Comprehensive and multi-granularity
  testing criteria for gauging the robustness of deep learning systems. arXiv
  preprint arXiv:1803.07519  (2018)

\bibitem{ma2018mode}
Ma, S., Liu, Y., Lee, W.C., Zhang, X., Grama, A.: Mode: automated neural
  network model debugging via state differential analysis and input selection.
  In: Proceedings of the 2018 26th ACM Joint Meeting on European Software
  Engineering Conference and Symposium on the Foundations of Software
  Engineering. pp. 175--186. ACM (2018)

\bibitem{ma2018characterizing}
Ma, X., Li, B., Wang, Y., Erfani, S.M., Wijewickrema, S., Schoenebeck, G.,
  Song, D., Houle, M.E., Bailey, J.: Characterizing adversarial subspaces using
  local intrinsic dimensionality. In: International Conference on Learning
  Representations (ICLR) (2018)

\bibitem{vanDerMaaten2008}
van~der Maaten, L., Hinton, G.: Visualizing data using {t-SNE}. Journal of
  Machine Learning Research  \textbf{9},  2579--2605 (2008),
  \url{http://www.jmlr.org/papers/v9/vandermaaten08a.html}

\bibitem{madry2018}
Madry, A., Makelov, A., Schmidt, L., Tsipras, D., Vladu, A.: Towards deep
  learning models resistant to adversarial attacks. In: International
  Conference on Learning Representations (ICLR) (2018)

\bibitem{madry2017towards}
Madry, A., Makelov, A., Schmidt, L., Tsipras, D., Vladu, A.: Towards deep
  learning models resistant to adversarial attacks. In: International
  Conference on Learning Representations (ICLR) (2018)

\bibitem{Mann47}
Mann, H.B., Whitney, D.R.: On a test of whether one of two random variables is
  stochastically larger than the other. Annals of Mathematical Statistics
  \textbf{18}(1),  50--60 (1947)

\bibitem{metricforadv}
Mao, C., Zhong, Z., Yang, J., Vondrick, C., Ray, B.: Metric learning for
  adversarial robustness. In: Advances in Neural Information Processing
  Systems. pp. 478--489 (2019)

\bibitem{metzen2017detecting}
Metzen, J.H., Genewein, T., Fischer, V., Bischoff, B.: On detecting adversarial
  perturbations. In: International Conference on Learning Representations
  (ICLR) (2017)

\bibitem{mirman2018differentiable}
Mirman, M., Gehr, T., Vechev, M.: Differentiable abstract interpretation for
  provably robust neural networks. In: International Conference on Machine
  Learning. pp. 3575--3583 (2018)

\bibitem{pmlr-v97-moon19a}
Moon, S., An, G., Song, H.O.: Parsimonious black-box adversarial attacks via
  efficient combinatorial optimization. In: Chaudhuri, K., Salakhutdinov, R.
  (eds.) Proceedings of the 36th International Conference on Machine Learning.
  Proceedings of Machine Learning Research, vol.~97, pp. 4636--4645. PMLR, Long
  Beach, California, USA (09--15 Jun 2019),
  \url{http://proceedings.mlr.press/v97/moon19a.html}

\bibitem{Ozdag2019OnTS}
Ozdag, M., Raj, S., Fernandes, S., Velasquez, A., Pullum, L., Jha, S.K.: On the
  susceptibility of deep neural networks to natural perturbations. In:
  AISafety@IJCAI (2019)

\bibitem{papernot2016limitations}
Papernot, N., McDaniel, P., Jha, S., Fredrikson, M., Celik, Z.B., Swami, A.:
  The limitations of deep learning in adversarial settings. In: 2016 IEEE
  European Symposium on Security and Privacy (EuroS\&P). pp. 372--387. IEEE
  (2016)

\bibitem{papernot2016distillation}
Papernot, N., McDaniel, P., Wu, X., Jha, S., Swami, A.: Distillation as a
  defense to adversarial perturbations against deep neural networks. In:
  Security and Privacy (SP), 2016 IEEE Symposium on. pp. 582--597. IEEE (2016)

\bibitem{pei2017deepxplore}
Pei, K., Cao, Y., Yang, J., Jana, S.: Deepxplore: Automated whitebox testing of
  deep learning systems. In: Proceedings of the 26th Symposium on Operating
  Systems Principles. pp. 1--18. ACM (2017)

\bibitem{pei2017towards}
Pei, K., Cao, Y., Yang, J., Jana, S.: Towards practical verification of machine
  learning: The case of computer vision systems. arXiv preprint
  arXiv:1712.01785  (2017)

\bibitem{CRADLE:ICSE2019}
Pham, H.V., Lutellier, T., Qi, W., Tan, L.: Cradle: Cross-backend validation to
  detect and localize bugs in deep learning libraries. In: Proceedings of the
  41st International Conference on Software Engineering. p. 1027–1038. ICSE
  ’19, IEEE Press (2019). \doi{10.1109/ICSE.2019.00107},
  \url{https://doi.org/10.1109/ICSE.2019.00107}

\bibitem{Qiu2020Quantifying}
Qiu, X., Meyerson, E., Miikkulainen, R.: Quantifying point-prediction
  uncertainty in neural networks via residual estimation with an i/o kernel.
  In: International Conference on Learning Representations (2020),
  \url{https://openreview.net/forum?id=rkxNh1Stvr}

\bibitem{Sawilowsky09}
Sawilowsky, S.: New effect size rules of thumb. Journal of Modern Applied
  Statistical Methods  \textbf{8},  597--599 (11 2009).
  \doi{10.22237/jmasm/1257035100}

\bibitem{automold}
Saxena, U.: Automold.
  \url{https://github.com/UjjwalSaxena/Automold--Road-Augmentation-Library/}

\bibitem{SenETAL05CUTE}
Sen, K., Marinov, D., Agha, G.: {CUTE}: A concolic unit testing engine for {C}.
  In: FSE (2005)

\bibitem{seshia2018formal}
Seshia, S.A., Desai, A., Dreossi, T., Fremont, D.J., Ghosh, S., Kim, E.,
  Shivakumar, S., Vazquez-Chanlatte, M., Yue, X.: Formal specification for deep
  neural networks. In: International Symposium on Automated Technology for
  Verification and Analysis. pp. 20--34. Springer (2018)

\bibitem{shaham2015understanding}
Shaham, U., Yamada, Y., Negahban, S.: Understanding adversarial training:
  Increasing local stability of neural nets through robust optimization. arXiv
  preprint arXiv:1511.05432  (2015)

\bibitem{NatPerV}
Shankar, V., Dave, A., Roelofs, R., Ramanan, D., Recht, B., Schmidt, L.: A
  systematic framework for natural perturbations from videos (06 2019)

\bibitem{44806}
Silver, D., Huang, A., Maddison, C.J., Guez, A., Sifre, L., van~den Driessche,
  G., Schrittwieser, J., Antonoglou, I., Panneershelvam, V., Lanctot, M.,
  Dieleman, S., Grewe, D., Nham, J., Kalchbrenner, N., Sutskever, I.,
  Lillicrap, T., Leach, M., Kavukcuoglu, K., Graepel, T., Hassabis, D.:
  Mastering the game of go with deep neural networks and tree search. Nature
  \textbf{529},  484--503 (2016),
  \url{http://www.nature.com/nature/journal/v529/n7587/full/nature16961.html}

\bibitem{simonyan2014very}
Simonyan, K., Zisserman, A.: Very deep convolutional networks for large-scale
  image recognition. In: International Conference on Learning Representations
  (ICLR) (2015)

\bibitem{simpson1949}
SIMPSON, E.H.: Measurement of diversity. Nature  \textbf{163}(4148),  688--688
  (1949), \url{https://doi.org/10.1038/163688a0}

\bibitem{2020-icse-misbehaviour-prediction}
Stocco, A., Weiss, M., Calzana, M., Tonella, P.: Misbehaviour prediction for
  autonomous driving systems. In: Proceedings of 42nd International Conference
  on Software Engineering. p. 12 pages. ICSE '20, ACM (2020)

\bibitem{Misbehaviour2020}
Stocco, A., Weiss, M., Calzana, M., Tonella, P.: Misbehaviour prediction for
  autonomous driving systems. In: International Conference of Software
  Engineering (ICSE) (2020)

\bibitem{sun2018concolic}
Sun, Y., Wu, M., Ruan, W., Huang, X., Kwiatkowska, M., Kroening, D.: Concolic
  testing for deep neural networks  (2018)

\bibitem{szegedy2013intriguing}
Szegedy, C., Zaremba, W., Sutskever, I., Bruna, J., Erhan, D., Goodfellow, I.,
  Fergus, R.: Intriguing properties of neural networks. In: International
  Conference on Learning Representations (ICLR) (2014)

\bibitem{pmlr-v80-teye18a}
Teye, M., Azizpour, H., Smith, K.: {B}ayesian uncertainty estimation for batch
  normalized deep networks. In: Dy, J., Krause, A. (eds.) Proceedings of the
  35th International Conference on Machine Learning. Proceedings of Machine
  Learning Research, vol.~80, pp. 4907--4916. PMLR, Stockholmsmässan,
  Stockholm Sweden (10--15 Jul 2018),
  \url{http://proceedings.mlr.press/v80/teye18a.html}

\bibitem{tian2017deeptest}
Tian, Y., Pei, K., Jana, S., Ray, B.: Deeptest: Automated testing of
  deep-neural-network-driven autonomous cars. In: International Conference of
  Software Engineering (ICSE), 2018 IEEE conference on. IEEE (2018)

\bibitem{tian2019deepinspect}
Tian, Y., Zhong, Z., Ordonez, V., Kaiser, G., Ray, B.: Testing dnn image
  classifier for confusion \& bias errors. In: International Conference of
  Software Engineering (ICSE) (2020)

\bibitem{tramer2017ensemble}
Tram{\`e}r, F., Kurakin, A., Papernot, N., Goodfellow, I., Boneh, D., McDaniel,
  P.: Ensemble adversarial training: Attacks and defenses. arXiv preprint
  arXiv:1705.07204  (2017)

\bibitem{Tsipras2019}
Tsipras, D., Santurkar, S., Engstrom, L., Turner, A., Madry, A.: Robustness may
  be at odds with accuracy. In: International Conference on Learning
  Representations (ICLR) (2019)

\bibitem{udacity-simulator}
{Udacity}: {A self-driving car simulator built with Unity}.
  \url{https://github.com/udacity/self-driving-car-sim} (2017), online;
  accessed 18 August 2019

\bibitem{Callisto}
{Udeshi}, S., {Jiang}, X., {Chattopadhyay}, S.: Callisto: Entropy-based test
  generation and data quality assessment for machine learning systems. In: 2020
  IEEE 13th International Conference on Software Testing, Validation and
  Verification (ICST). pp. 448--453 (2020)

\bibitem{wang2017residual}
Wang, F., Jiang, M., Qian, C., Yang, S., Li, C., Zhang, H., Wang, X., Tang, X.:
  Residual attention network for image classification. In: Proceedings of the
  IEEE Conference on Computer Vision and Pattern Recognition. pp. 3156--3164
  (2017)

\bibitem{10.1109/ICSE.2019.00126}
Wang, J., Dong, G., Sun, J., Wang, X., Zhang, P.: Adversarial sample detection
  for deep neural network through model mutation testing. In: Proceedings of
  the 41st International Conference on Software Engineering. p. 1245–1256.
  ICSE '19, IEEE Press (2019). \doi{10.1109/ICSE.2019.00126},
  \url{https://doi.org/10.1109/ICSE.2019.00126}

\bibitem{wang2018mixtrain}
Wang, S., Chen, Y., Abdou, A., Jana, S.: Mixtrain: Scalable training of
  formally robust neural networks. arXiv preprint arXiv:1811.02625  (2018)

\bibitem{WangFormal2018}
Wang, S., Pei, K., Whitehouse, J., Yang, J., Jana, S.: Efficient formal safety
  analysis of neural networks. In: Proceedings of the 32Nd International
  Conference on Neural Information Processing Systems. pp. 6369--6379. NIPS'18,
  Curran Associates Inc., USA (2018),
  \url{http://dl.acm.org/citation.cfm?id=3327345.3327533}

\bibitem{wang2018formal}
Wang, S., Pei, K., Whitehouse, J., Yang, J., Jana, S.: Formal security analysis
  of neural networks using symbolic intervals. USENIX Security Symposium
  (2018)

\bibitem{wong2018scaling}
Wong, E., Schmidt, F., Metzen, J.H., Kolter, J.Z.: Scaling provable adversarial
  defenses. In: Advances in Neural Information Processing Systems. pp.
  8400--8409 (2018)

\bibitem{xiao2018generating}
Xiao, C., Li, B., Zhu, J.Y., He, W., Liu, M., Song, D.: Generating adversarial
  examples with adversarial networks. In: 27th International Joint Conference
  on Artificial Intelligence (IJCAI) (2018)

\bibitem{spatial2018}
Xiao, C., Zhu, J.Y., Li, B., He, W., Liu, M., Song, D.: Spatially transformed
  adversarial examples. In: International Conference on Learning
  Representations (ICLR) (2018)

\bibitem{fmnist}
Xiao, H., Rasul, K., Vollgraf, R.: Fashion-mnist: a novel image dataset for
  benchmarking machine learning algorithms (2017)

\bibitem{trade_alp_spatial}
Yang, F., Wang, Z., Heinze-Deml, C.: Invariance-inducing regularization using
  worst-case transformations suffices to boost accuracy and spatial robustness.
  In: Advances in Neural Information Processing Systems 32. pp. 14757--14768
  (2019)

\bibitem{svhn}
Yuval~Netzer, T.W., Coates, A., Bissacco, A., Wu, B., Ng, A.Y.: Reading digits
  in natural images with unsupervised feature learning. In: NIPS Workshop on
  Deep Learning and Unsupervised Feature Learning (2011)

\bibitem{Zagoruyko2016WRN}
Zagoruyko, S., Komodakis, N.: Wide residual networks. In: BMVC (2016)

\bibitem{Apricot:ASE2019}
{Zhang}, H., {Chan}, W.K.: Apricot: A weight-adaptation approach to fixing deep
  learning models. In: 2019 34th IEEE/ACM International Conference on Automated
  Software Engineering (ASE). pp. 376--387 (Nov 2019).
  \doi{10.1109/ASE.2019.00043}

\bibitem{zhang2018deeproad}
Zhang, M., Zhang, Y., Zhang, L., Liu, C., Khurshid, S.: Deeproad: Gan-based
  metamorphic autonomous driving system testing. arXiv preprint
  arXiv:1802.02295  (2018)

\bibitem{zhao2017generating}
Zhao, Z., Dua, D., Singh, S.: Generating natural adversarial examples. In:
  International Conference on Learning Representations (ICLR) (2018)

\bibitem{DeepBillboard2020}
Zhou, H., Li, W., Kong, Z., Guo, J., Zhang, Y., Zhang, L., Yu, B., Liu, C.:
  Deepbillboard: Systematic physical-world testing of autonomous driving
  systems. In: International Conference of Software Engineering (ICSE) (2020)

\end{thebibliography}
